\RequirePackage[T1]{fontenc}
\documentclass[12pt]{article}
\pdfoutput=1 
\usepackage[height=8.85in,width=6.45in]{geometry}

\usepackage[utf8]{inputenc}
\usepackage{amsmath}
\usepackage{amssymb}
\usepackage{mathtools}
\numberwithin{equation}{section}
\usepackage{slashed}
\usepackage{braket}
\usepackage[svgnames]{xcolor}
\usepackage[colorlinks,linktocpage=true,citecolor=DarkGreen,linkcolor=FireBrick]{hyperref}
\usepackage{cite}
\usepackage{graphicx}
\usepackage{tikz}
\usepackage{tikz-cd}
\usepackage{times}
\usepackage[scaled]{couriers}
\usepackage{bm}
\usepackage{subfig}
\usepackage{mathrsfs}
\usepackage{amsthm}

\usepackage{xcolor}
\usepackage{mdframed}

\renewenvironment{figure}[1][]{
  \begin{originalfigure}[#1]
    \begin{mdframed}[linecolor=black!0,backgroundcolor=black!1]
}{
    \end{mdframed}
  \end{originalfigure}
}

\theoremstyle{plain}

\theoremstyle{definition}

\numberwithin{thm}{section}

\def\d{{\rm d}}

\DeclareMathOperator{\tr}{tr}

\usepackage{slashed}

\def\a{\alpha}

\def\cH{{\cal H}}

\def\cL{{\cal L}}

\def\cO{{\cal O}}

\def\cR{{\cal R}}

\def\bR{{\mathbb R}}

\def\bZ{{\mathbb Z}}

\def\sC{{\mathsf C}}

\def\si{{\mathsf i}}

\def\so{{\mathsf o}}

\def\su{{\mathsf u}}

\def\SO{\mathrm{SO}}

\def\Spin{\mathrm{Spin}}

\def\su{\mathfrak{su}}

\def\so{\mathfrak{so}}

\def\e{\mathfrak{e}}
\def\g{\mathfrak{g}}

\def\beq#1\eeq{\begin{align}#1\end{align}}

\newcommand{\diff}{\mathrm{d}}

\newcommand{\hetE}{(E_8\times E_8)\rtimes \bZ_2}
\newcommand{\hetS}{\Spin(32)/\bZ_2}

\newcounter{myfigure}[figure]
\newcounter{pseudofigure}
\renewcommand{\themyfigure}{\thepseudofigure.\alph{myfigure}}
\newcommand{\mycaption}[1]{\setcounter{pseudofigure}{\thefigure}\addtocounter{pseudofigure}{1}\refstepcounter{myfigure}Figure~{\themyfigure}: #1}

\begin{document}

\begin{titlepage}

\begin{flushright}
TU-1247
\end{flushright}

\vskip 3cm

\begin{center}

{\Large \bfseries Black $p$-Branes in Heterotic String Theory}

\vskip 1cm
Masaki Fukuda$^{1}$, Shun K. Kobayashi$^{1}$, Kento Watanabe$^{1,2}$, and Kazuya Yonekura$^{1}$
\vskip 1cm

\begin{tabular}{ll}
$^1$ Department of Physics, Tohoku University, Sendai 980-8578, Japan \\
$^2$ Proxima Technology Inc., KDX Okachimachi Bld., 5-24-16, Ueno, Tokyo, 110-0005, Japan
\end{tabular}

\vskip 1cm

\end{center}

\noindent

Heterotic string theory has nonsupersymmetric branes whose existence is suggested by the cobordism conjecture. 
We numerically construct static, spherically symmetric, and asymptotically flat black brane solutions in ten-dimensional heterotic superstring theories for 0- and 4-branes. 
These branes carry charges that are measured by Chern classes on the sphere surrounding the branes. For the extremal case, the solutions have a throat region with a linear dilaton profile as expected from the corresponding world-sheet theory. We also construct non-extremal solutions by compactifying the time direction. 
To verify the reliability of our numerical calculations, we confirm that they reproduce the known analytical solutions for the 6-brane.
Our black brane solutions provide evidence supporting the existence of such branes in heterotic string theory.

\end{titlepage}

\setcounter{tocdepth}{3}

\tableofcontents

\section{Introduction}

In string theory, there are various types of branes, some of which can be predicted by the completeness hypothesis~\cite{Polchinski:2003bq,Banks:2010zn}. According to this hypothesis, for each possible charge in quantum gravity, there must exist objects carrying them. The most well-known example of such objects are D-branes. If we consider RR-charges, we can predict the existence of objects carrying these charges based on the hypothesis. As shown by Polchinski~\cite{Polchinski:1995mt}, D-branes indeed have RR-charges. The completeness hypothesis is 
based on the absence of global symmetries in quantum gravity (e.g. \cite{Misner:1957mt,Banks:1988yz,Garfinkle:1990qj,Polchinski:2003bq,Banks:2010zn,Harlow:2018jwu,Harlow:2018tng}), which in turn is partly based on the idea that black hole evaporation through Hawking radiation would contradict global charge conservation.

There are also somewhat subtler kinds of charges. They are given by any topologically nontrivial field configuration at spatial infinity. The cobordism conjecture~\cite{McNamara:2019rup} predicts that there exist physical configurations that realize any given topologically nontrivial configuration at spatial infinity. Some of these configurations are localized objects which can be regarded as branes (see Section~2 and 3 of \cite{Kaidi:2024cbx} for detailed explanations).

In this paper, we consider the ten-dimensional heterotic superstring theories with gauge group $G = \hetE~\text{or}~\hetS$. In heterotic string theory, there are charges characterized by topologically nontrivial gauge field configurations of $G$ at spatial infinity. Suppose we are interested in a $p$-brane. Its codimension is $9-p$, and the sphere at infinity surrounding the brane is $S^{n}$, where $n=8-p$. We can consider topologically nontrivial gauge field configurations on the $S^n$. These configurations are known to be classified by the homotopy group $\pi_{n-1}(G)$. 
By applying the cobordism conjecture to these charges, the existence of $p$-branes with $p=0,4,6,$ and 7 in heterotic string theory was suggested\cite{Kaidi:2023tqo,Kaidi:2024cbx}. (For related work, see also \cite{Hellerman:2006ff,Hellerman:2007zz,Kaidi:2020jla,BoyleSmith:2023xkd,Fraiman:2023cpa,DeFreitas:2024yzr,Yonekura:2024spl,Etheredge:2024amg,Hamada:2024cdd}.)

Among the branes studied in \cite{Kaidi:2023tqo,Kaidi:2024cbx}, the $p$-branes with $p=6$ and $7$ are new objects. In contrast, the 0-brane was proposed as an object on which heterotic strings can end \cite{Polchinski:2005bg}, while the 4-brane was introduced as a brane in the context of M-theory \cite{Bergshoeff:2006bs}. 
The relations to the cobordism conjecture are further discussed in \cite{Basile:2023knk,Debray:2023rlx,Kneissl:2024zox}. 
(See also e.g. \cite{Montero:2020icj,Blumenhagen:2021nmi,Blumenhagen:2022mqw,Andriot:2022mri,Velazquez:2022eco,Dierigl:2020lai,Dierigl:2022reg,Debray:2023yrs,Dierigl:2023jdp,Alvarez-Garcia:2024vnr,Angius:2024pqk} for other work related to the cobordism conjecture.)

The purpose of this paper is to construct black $p$-brane solutions for the 0-brane and 4-brane, providing further evidence for the existence of these branes.
Apart from a subtle choice of charge, the black brane solution corresponding to the 6-brane was essentially known~\cite{Garfinkle:1990qj,Horowitz:1991cd} (which is written down in \cite{Kaidi:2023tqo,Kaidi:2024cbx} in a modern normalization of fields). 
On the other hand, although the black brane solution for the 4-brane was discussed in \cite{Bergshoeff:2006bs}, a solution connecting the near-horizon region and asymptotic infinity was not provided. Furthermore, black brane solutions for the cases $p=0$ and $7$ were not addressed in these papers.\footnote{
We will not try to construct black brane solutions for the 7-brane, since the 7-brane involves tachyon condensation~\cite{Kaidi:2023tqo,Kaidi:2024cbx}. 
} The exact worldsheet theories for the throat regions of the branes are constructed in \cite{Kaidi:2023tqo,Kaidi:2024cbx}, but the whole solutions including the asymptotically flat region were missing.\footnote{Black brane solutions for the 4-brane were also studied in \cite{Abbott:2008cd}, but our conclusions are different. }
As in the case of the various black brane solutions constructed in \cite{Horowitz:1991cd}, 
black brane solutions would similarly provide evidence for the existence of the branes in \cite{Polchinski:2005bg,Bergshoeff:2006bs,Kaidi:2023tqo,Kaidi:2024cbx}.

In general, equations of motion are second-order nonlinear differential equations. Unless we are lucky or a configuration is supersymmetric, it is not easy to find an analytical solution. Therefore, we solve the equations numerically to find solutions with the desired properties.

\begin{figure}[htb]
  \begin{minipage}{0.48\columnwidth}
    \begin{center}
      \begin{tikzpicture}
        \draw [thick] (0,0)--(2,1)--(2,6)--(0,5)--(0,0)--cycle;
        \draw [thick] (-4,2+1/4).. controls (-2,2+1/4) and (-1,2+1/4) ..(0,3/2+1/4);
        \draw [dashed] (0,3/2+1/4).. controls (1/2,5/4+1/4) ..(1,5/4-1/16);
        \draw [thick] (-4,3).. controls (-2,3) and (-1,3) ..(0,7/2);
        \draw [dashed] (0,7/2).. controls (1/2,4-1/4) ..(1,4+1/16);
        \draw (1,7/2+1/4).. controls (1/2,7/2-1/12) and (1/2,7/4+1/12) ..(1,7/4-1/4);
        \draw (1,7/2).. controls (3/2,7/2-1/16) and (3/2,7/4+1/16) ..(1,7/4);
      \end{tikzpicture}
      \mycaption{extremal case}
      \label{fig:extremal}
    \end{center}
  \end{minipage}
  \begin{minipage}{0.02\columnwidth}
    \hspace{0.01\columnwidth}
  \end{minipage}
  \begin{minipage}{0.48\columnwidth}
    \begin{center}
      \begin{tikzpicture}
        \draw [thick] (0,0)--(2,1)--(2,6)--(0,5)--(0,0)--cycle;
        \draw [thick] (-4,2+1/4).. controls (-2,2+1/4) and (-1,2+1/4) ..(0,3/2+1/4);
        \draw [dashed] (0,3/2+1/4).. controls (1/2,5/4+1/4) ..(1,5/4-1/16);
        \draw [thick] (-4,3).. controls (-2,3) and (-1,3) ..(0,7/2);
        \draw [dashed] (0,7/2).. controls (1/2,4-1/4) ..(1,4+1/16);
        \draw [thick] (-4,3).. controls (-5,3) and (-5,2+1/4) ..(-4,2+1/4);
        \draw [] (1,7/2+1/4).. controls (1/2,7/2-1/12) and (1/2,7/4+1/12) ..(1,7/4-1/4);
        \draw [] (1,7/2).. controls (3/2,7/2-1/16) and (3/2,7/4+1/16) ..(1,7/4);
      \end{tikzpicture}
      \mycaption{non-extremal case}
      \label{fig:non-extremal}
    \end{center}
  \end{minipage}
  \caption{Left (\ref{fig:extremal}): the extremal case. Right (\ref{fig:non-extremal}): the non-extremal case.
The right figure is only schematic because the $S^1$ for the Euclidean time direction and the $S^n$ for the sphere surrounding the brane are not clearly distinguished in the figure.
   In both cases, the supergravity solutions have an asymptotically flat region. The extremal case has a throat region with a linear dilaton, while the non-extremal case terminates at a horizon where the $S^1$ shrinks to zero. For the extremal case, the horizon is located at the infinity of the throat.}
\end{figure}

The structure of this paper is as follows. In Section \ref{sec:extremal}, we investigate extremal solutions. In this case, the supergravity solution is expected to have a throat region with linear dilaton, and also an asymptotically flat region, as illustrated in Figure~\ref{fig:extremal}. 
After briefly discussing gauge field configurations on $S^n$, we reduce the supergravity action to a one-dimensional system by using spherical symmetry.
The asymptotic behavior in the throat region can be studied analytically, which is used as an initial condition for the one-dimensional differential equations.
Then we provide numerical solutions for these equations and confirm that they possess the properties shown in Figure~\ref{fig:extremal}.\footnote{Our numerical calculation method is somewhat similar to that used in \cite{Aharony:2019zsx}.}
To verify the reliability of our numerical calculations, 
we perform the same calculations for the 6-brane case and compare them with the known analytical solution mentioned above. 

In Section \ref{sec:nonextremal}, we extend the analysis to the non-extremal case. We use Euclidean signature and compactify the Euclidian time direction on $S^1$ to consider a finite temperature system. Then, the supergravity solution is expected to have a horizon 
where the radius of $S^1$ shrinks smoothly to zero. The situation is schematically illustrated in Figure~\ref{fig:non-extremal}. 
As in the extremal case, we reduce the supergravity to a one-dimensional system, find the asymptotic solution near the horizon, and use it as an initial condition for the numerical solution. We also provide analytical solutions for the limit in which we only have the throat region and the horizon, as in the cigar geometry in \cite{Witten:1991yr}.

In Appendix~\ref{sec:gaugefield}, we discuss when and how supergravity solutions are reliable in the sense that $\alpha'$ corrections are negligible, and give some examples of  gauge field configurations.
Finally, in Appendix \ref{app:n9sol}, we give an analytical solution for the case $n=9$, although physical meaning of the solution is not yet clear.

\section{Extremal solutions}\label{sec:extremal}

In this section, we study extremal solutions in which the worldvolume of a $p$-brane has the $\SO(p,1)$ Lorentz symmetry. In the extremal case, the solution is expected to have a throat region and an asymptotic flat region as illustrated in Fig. \ref{fig:extremal}. This case is a limit of the more general non-extremal solutions studied in Section~\ref{sec:nonextremal}, but it may be illustrative to study this case explicitly.

\subsection{Gauge field configurations}\label{sec:extremal:gauge}

Let us first set some notations. For a 2-form gauge field strength $F$, we denote $ |F|^2 := \frac{1}{2}F_{\mu_1\mu_2}^\dagger F^{\mu_1\mu_2}$.
We define $\tr'$ for nonabelian gauge algebras $\so(32)$ and $\e_8 \times \e_8$ as
\beq
\tr' = \left\{ 
\begin{array}{ll}
\frac12 \tr_{\rm fundamental} & \so(32)  \\
\frac{1}{60} \tr_{\rm adjoint} & \e_8 \times \e_8
\end{array}
\right.\label{eq:tr}
\eeq
where $\tr_{\rm fundamental}$ and $ \tr_{\rm adjoint}$ are the traces in the fundamental and the adjoint representations, respectively.
This normalization is chosen by the following reason. If we consider a $\su(2)$ subalgebra corresponding to a simple root of $\so(32)$ or $\e_8 \times \e_8$ and put a gauge field $F_{\su(2)}$ in that subalgebra, we have $\tr' F^2 = \tr F_{\su(2)}^2$ where $\tr$ for $\su(2)$ is the trace in the fundamental representation. 

The relevant part of the supergravity action is given by
\begin{equation}
  S_{(10)} \propto \int\diff^{10}x\ \sqrt{-G}e^{-2\Phi}\left(\mathcal R_{(G)} + 4\left(\nabla^{(G)}\Phi\right)^2 - \frac{\alpha'}{2}\tr' \left(|F|^2\right) + \cdots\right),
\end{equation}
where $G_{\mu\nu}$ is the metric, $\cR_G$ is the Ricci scalar, $\Phi$ is the dilaton, and $\nabla^{(G)}$ is the covariant derivative with respect to $G$. We have omitted the $B$-field since it is not involved in our solutions.

We will construct black $p$-brane solutions with certain gauge field configurations.
The brane has codimension $ 10-(p+1)$ which we denote as $n+1$, i.e., $n=8-p$. We use polar coordinates for these directions. We denote the radial direction and the sphere describing the angular directions as $\bR_{\rm radial}$ and $S^{n}$, respectively.

In the construction of supergravity solutions, the only information that is practically necessary is that the gauge configuration has a spherical symmetry, and $\tr' |F|^2$ is constant times $R^{-4}$ where $R$ is the radius of $S^n$. We can consider any gauge field configuration as far as this condition is satisfied. Thus we simply assume
\beq
\tr' |F|^2 = \frac{\sC}{R^4} \label{eq:Fsquared}
\eeq
where $\sC>0$ is a constant. More explicit gauge field configurations are discussed in Appendix~\ref{sec:gaugefield}.

\subsection{Supergravity equations}\label{sec:extremal:equation}

For the purpose of finding supergravity solutions, it is convenient to recall the following point about heterotic string theories. On the worldsheet of a heterotic string theory, we have the fields $X^M$ ($M=0,1,\cdots,9)$ describing the target space coordinates as well as their superpartners. When we consider an extremal black brane solution, the directions tangent to the brane and the directions normal to the brane can be decoupled in the worldsheet action. This is achieved by taking the target space metric to be a product form. We can take the tangent directions to be just free fields. 

In supergravity, the above worldsheet situation corresponds to the following string frame metric. Let $X^\alpha$ ($\alpha=0,\cdots,p$) be the coordinates for the tangent directions. Then the string frame metric is of the form
\beq
\d s^2_{(10)} = \d X^\alpha \d X_\alpha +  \d s^2_{(n+1)} \label{eq:extremalansatz}
\eeq
where $ \d s^2_{(n+1)}$ is the metric for the normal directions. This metric has $\so(p,1)$ Lorentz symmetry acting on $X^\alpha$ and hence it will correspond to an extremal black brane solution. We will review more direct justification of this ansatz for the metric in the paragraph containing \eqref{eq:Sigmaeq}.

In the above ansatz, the tangent directions $X^\alpha$ play no role and hence we can simply neglect them. Then we can regard the spacetime to be just an $(n+1)$-dimensional space. We take its metric to be
\beq
 \d s^2_{(n+1)} = N^2 \d r^2 + R^2 \d \Omega_n^2, \label{eq:extmetric}
\eeq
where $ \d \Omega_n^2$ is the metric of the round sphere $S^n$ with unit radius, $r$ is a coordinate for the radial direction $\bR_{\rm radial}$, and $N=N(r)$ and $R=R(r)$ are functions of $r$. 

It might be convenient to see the above metric \eqref{eq:extmetric} as a ``Friedmann-Lemaitre-Robertson-Walker (FLRW) metric'' for an expanding universe, but with a Euclidean signature time $r$. The $R$ is the ``scale factor'' of the universe, $ \d \Omega_n^2$ is the metric for the ``positive curvature space'', and $N$ is the ``lapse function''. 
The Ricci scalar of the FLRW metric with Euclidean signature ``time'' $r$ is given by\footnote{It is convenient to first compute the Ricci scalar in the gauge $N=1$, and then go to more general gauge by noting that $N \d r$ is the gauge invariant combination.}
\beq
\cR_{G} = -2n   \frac{1}{N} \frac{\d}{ \d r} \left( \frac{1}{N} \frac{\d \log R}{ \d r}   \right) - n(n+1) \left( \frac{1}{N}  \frac{\d \log R}{ \d r}  \right)^2 + n(n-1)R^{-2}. \label{eq:ExtRs}
\eeq
By using it, the action becomes
\beq
S_{(n+1)} &\propto V_{p+1} \int \d^{(n+1)} x  \sqrt{G} e^{-2\Phi}\left(\mathcal R_{G} + 4\left(\nabla^{G}\Phi\right)^2 - \frac{\alpha'}{2}\tr' \left(|F|^2\right)\right)   \nonumber \\
& \propto \int \d r  N R^n e^{-2\Phi} \Bigg( 
-2n   \frac{1}{N} \frac{\d}{ \d r} \left( \frac{1}{N} \frac{\d \log R}{ \d r}   \right) - n(n+1) \left( \frac{1}{N}  \frac{\d \log R}{ \d r}  \right)^2  \nonumber \\
& \qquad \qquad  \qquad \qquad \qquad + n(n-1)R^{-2} + 4\left( \frac{1}{N} \frac{ \d \Phi}{\d r} \right)^2 - \frac{\alpha' \sC }{2R^4}  \Bigg)\label{eq:2_13}
\eeq
where $V_{p+1}$ is the constant volume of the $(p+1)$-directions which we neglect, and we have used \eqref{eq:Fsquared}. 

For later convenience, we define
\beq
\ell_0 =  \sqrt{\frac{\alpha' \sC}{n(n-1)}}  , \label{eq:Sn-radius}
\eeq
and change variables as
\beq
r =  \sqrt{\frac{8}{n(n-1)} } \, \ell_0\tau , \qquad R = \ell_0   \,  e^\sigma , \qquad \Phi  =\phi + \frac{n}{2} \sigma.  \label{eq:variablechange}
\eeq
Physical meanings of $\sigma$ and $\phi$ are as follows. We can think that the $S^n$ is a compactified space. Then $\sigma$ is a scalar field obtained by the compactification, and $\phi$ is the effective dilaton after the compactification. In terms of these variables, the metric $ \d s^2_{(n+1)} $ is given by
\beq
 \d s^2_{(n+1)} /\ell_0^2 = \frac{8}{n(n-1)} N^2 \d \tau^2 + e^{2\sigma} \d \Omega_n^2.
\eeq

Denoting derivatives with respect to $\tau$ by primes such as $\frac{\d }{\d \tau} \phi= \phi'$, we get (after dropping total derivative terms)
\beq
S_{(n+1)} \propto \int \d \tau  \cL,
\eeq
where
\beq
\cL  =  N e^{- 2\phi} \left( - \frac{n}{8} (N^{-1} \sigma')^2 + \frac{1}{2} (N^{-1} \phi ')^2 +   e^{-2\sigma}  -  \frac{1}{2} e^{-4\sigma} \right). \label{eq:extremalLagrangian}
\eeq
Equations of motion can be derived by using this Lagrangian.

Although the action obtained above is enough, it is also possible to go to ``canonical formalism'' by regarding $\tau$ as a ``time''. For this purpose, we define the ``canonical momenta'' $\Pi_\sigma$ and $\Pi_\phi$ as
\beq
\Pi_\sigma &=  \frac{\partial \cL}{\partial \sigma'} = -\frac{n}{4} e^{-2\phi} (N^{-1} \sigma'),  \nonumber \\
\Pi_\phi &=  \frac{\partial \cL}{\partial \phi'} =  e^{-2\phi} (N^{-1} \phi')    . \label{eq:CanMom}
\eeq
Then we define the ``Hamiltonian density'' $\cH$ by
\beq
 \cH = \left( \Pi_{\sigma}   \sigma' + \Pi_{\phi}    \phi' - \cL \right) = N H
 \eeq
 where
 \beq
H=  e^{2\phi} \left( - \frac{2}{n}\Pi_\sigma^2 + \frac{1}{2} \Pi_\phi^2 \right) +e^{-2\phi} \left(- e^{-2\sigma}  + \frac{1}{2} e^{-4\sigma} \right)   .
\eeq
The equations of motion for $\sigma$ and $\phi$ are
\beq
N^{-1} \sigma' = \frac{\partial H}{\partial \Pi_\sigma}, \quad N^{-1}\phi' = \frac{\partial H}{\partial \Pi_\phi}, \quad 
N^{-1} \Pi'_\sigma = - \frac{\partial H}{\partial \sigma}, \quad N^{-1} \Pi'_\phi = - \frac{\partial H}{\partial \phi}. \label{eq:eoms}
\eeq
In addition to these equations, we have the equation of motion that comes from $N$. It gives the ``Hamiltonian constraint''
\beq
H=0.
\eeq
The equations of motion for $\sigma, \phi, \Pi_\sigma, \Pi_\phi$ guarantees that $H$ is conserved. The Hamiltonian constraint says that the conserved value must be exactly zero.

The function $N$ can be taken freely (under reasonable conditions such as $N>0$) since we can perform gauge transformations or diffeomorphisms $\tau \to f(\tau)$ for any reasonable function $f$. In the gauge $N=1$, the equations of motion and the Hamiltonian constraint are explicitly given by
\beq
0 &= \frac{n}{4} \left(\sigma'' -2\phi' \sigma'  \right) -2( e^{-2\sigma} -e^{-4\sigma} ), \label{eq:eomsigma}\\
0&= \phi'' -\phi'^2    - \frac{n}{4}  \sigma'^2 + 2e^{-2\sigma}  -  e^{-4\sigma}, \label{eq:eomphi}\\
0&=  \frac{n}{4}   \sigma'^2 -    \phi '^2 +  2e^{-2\sigma}  -  e^{-4\sigma} \label{eq:hamiltonianconst} ,
\eeq
where the first two equations are the equations of motion and the last one is the Hamiltonian constraint. 

\subsection{Qualitative features of the extremal solution}\label{sec:extremal:qual}
Let us study some qualitative features. 
First, the flat region which is far away from the brane is given by $\Phi=\textrm{const.}$ (say $\Phi=0$) and $R=r$ in the gauge $N=1$, and hence
\beq
\sigma = \log \tau + \frac12 \log \left( \frac{8}{n(n-1)}  \right), \quad \phi = -\frac{n}{2} \sigma, \qquad (N=1) \label{eq:flatosol}
\eeq
where we have used \eqref{eq:variablechange}. One can check that this is indeed a solution in the region $\tau \to \infty$ where the term proportional to $e^{-4\sigma}$ is neglected.

The limit $\tau \to -\infty$ will be given by a throat region, in which $\sigma$ and $N^{-1} \phi'$ are constant. In the gauge $N=1$, it is  explicitly given by
\beq
\sigma =0, \qquad   \phi = - \tau +\text{(const.)}, \qquad (N=1) \label{eq:throatsol}
\eeq
where we have chosen the sign of $\phi'$ so that $\phi$ increases for decreasing $\tau$. One can check that this is a solution of the equations. It is possible to show that $\phi'$ must be negative for any $\tau\in\mathbb R$ by the following argument. Eliminating $\phi'$ and $\sigma$ from \eqref{eq:eomphi} by using \eqref{eq:hamiltonianconst}, we obtain $\phi''=\frac{n}{2}\sigma'^2\geq 0$ and hence $\phi'$ is monotonically increasing. We also have $\phi'\to 0$ in the flat region $\tau\to\infty$. Thus, $\phi'$ must approach zero from the negative side, and hence $\phi' <0$ in the entire region.

Let us study the throat region in a little more detail. The zeroth order solution is given by \eqref{eq:throatsol}. Consider small perturbations around it,
\beq
\sigma=0+\delta \sigma, \qquad  \phi = - \tau +\text{(const.)} + \delta \phi.
\eeq
Let $\epsilon $ be an infinitesimal number. It will turn out that if $\delta \sigma =\cO(\epsilon)$, then $\delta \phi = \cO(\epsilon^2)$. We expand the equations of motion and retain leading terms in $\epsilon$. Then we get
\beq
0 &= \frac{n}{4} (\delta \sigma'' +2 \delta \sigma')  -4\delta \sigma, \\
0&= \delta \phi'' +2 \delta \phi'  -  \frac{n}{4} \delta \sigma'^2 - 4\delta \sigma^2   , \\
0&=  \frac{n}{4} \delta \sigma'^2+2 \delta\phi '  - 4\delta \sigma^2.
\eeq
Solutions of these equations are given as follows. We define
\beq
\alpha_{\pm} = -1 \pm  \sqrt{1+\frac{16}{n}} 
\eeq
Then, a general solution of the first equation is given by
\beq
\delta \sigma = A_+ e^{ \alpha_+ \tau } + A_- e^{ \alpha_- \tau}.
\eeq
where $A_\pm$ are integration constants. The third equation gives
\beq
\delta \phi = B + \frac{n}{8} \left( A_+^2   e^{ 2\alpha_+ \tau }+A_-^2  e^{ 2\alpha_- \tau }  \right) - 4 A_+ A_- e^{ -2 \tau }  .
\eeq
where $B$ is an integration constant. We want the $\delta \sigma$ to go to zero in the limit $\tau \to -\infty$, so we need to set 
$A_-=0$. Denoting $A=A_+$, the solution is
\beq
\sigma \simeq A e^{ \alpha_+ \tau }, \qquad \phi \simeq -\tau + B + \frac{n}{8} A^2  e^{ 2\alpha_+ \tau }. \label{eq:BoundaryCondition}
\eeq

\subsection{Numerical solutions : the extremal case}\label{sec:numerical:ext}

In this section, we give numerical solutions of the equations of motion \eqref{eq:eoms}. 
For a clearer physical interpretation, we choose the gauge $N=1$ and plot $\Phi$ and $\tilde R=e^{\sigma}=R/\ell_0$, 
where $\ell_0$ is the throat radius defined in \eqref{eq:Sn-radius}.

\begin{figure}[htb]
  \begin{center}
    $n=4$\\
    \begin{minipage}{0.48\columnwidth}
      \begin{center}
        \includegraphics[scale=0.74]{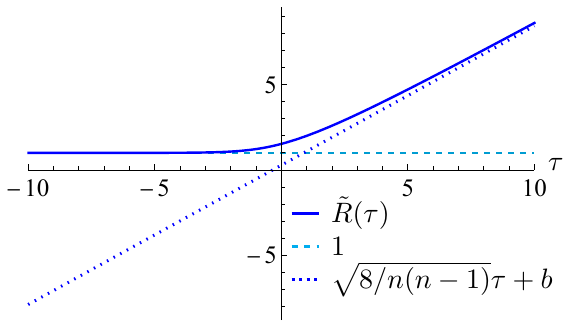}
      \end{center}
    \end{minipage}
    \begin{minipage}{0.02\columnwidth}
      \hspace{0.02\columnwidth}
    \end{minipage}
    \begin{minipage}{0.48\columnwidth}
      \begin{center}
        \includegraphics[scale=0.74]{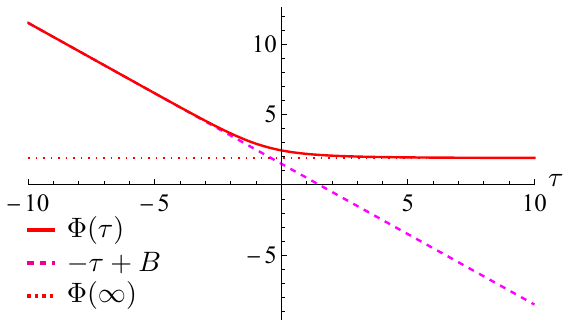}
      \end{center}
    \end{minipage}
  \end{center}
  \caption{The solution for $n=4$. Left: $\tilde R(\tau)$. Right: $\Phi(\tau)$.}
  \label{fig:4branesol}
\end{figure}

\begin{figure}[htb]
  \begin{center}
    $n=8$\\
    \begin{minipage}{0.48\columnwidth}
      \begin{center}
        \includegraphics[scale=0.74]{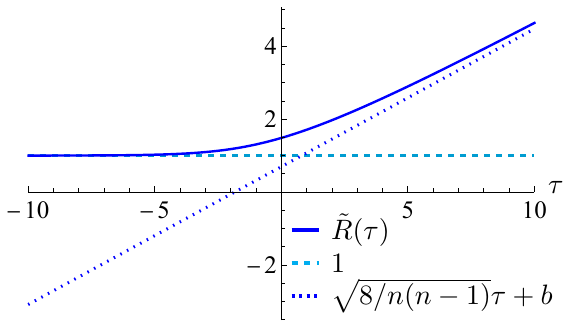}
      \end{center}
    \end{minipage}
    \begin{minipage}{0.02\columnwidth}
      \hspace{0.02\columnwidth}
    \end{minipage}
    \begin{minipage}{0.48\columnwidth}
      \begin{center}
        \includegraphics[scale=0.74]{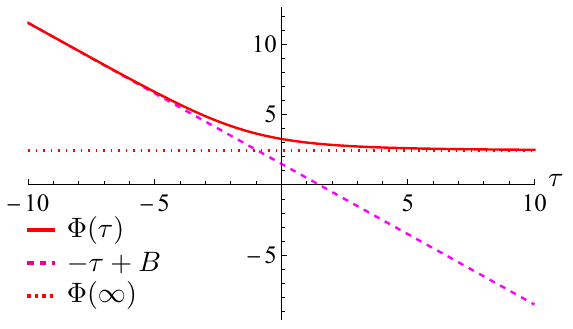}
      \end{center}
    \end{minipage}
  \end{center}
    \caption{The solution for $n=8$. Left: $\tilde R(\tau)$. Right: $\Phi(\tau)$.}
  \label{fig:8branesol}
\end{figure}

\begin{figure}[htb]
  \begin{center}
    \begin{minipage}{0.48\columnwidth}
      \begin{center}
        $n=4$
        \includegraphics[scale=0.74]{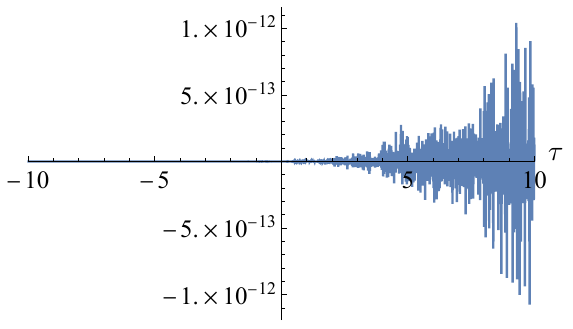}
      \end{center}
    \end{minipage}
    \begin{minipage}{0.02\columnwidth}
      \hspace{0.02\columnwidth}
    \end{minipage}
    \begin{minipage}{0.48\columnwidth}
      \begin{center}
        $n=8$
        \includegraphics[scale=0.74]{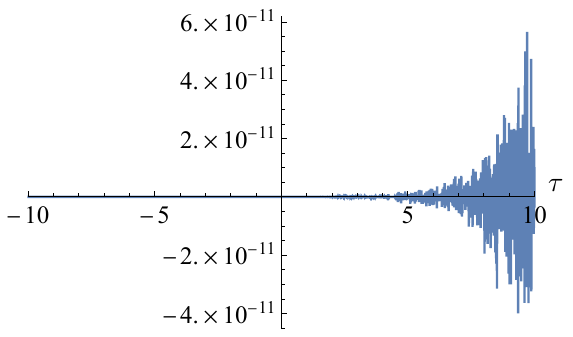}
      \end{center}
    \end{minipage}
    \caption{The Hamiltonian constraint \eqref{eq:hamiltonianconst}$\times e^{-2\phi}$. These values are zero within the limits of numerical accuracy. }
    \label{fig:04braneHconstraint}
  \end{center}
\end{figure}

Figures \ref{fig:4branesol} and \ref{fig:8branesol} show the results of the numerical solutions of the equations \eqref{eq:eomsigma} and \eqref{eq:eomphi} for $n=4$ and $n=8$, respectively. The horizontal axis in these figures corresponds to $\tau$. The calculations are performed from $\tau=-40$ in the direction of increasing $\tau$. Equation \eqref{eq:BoundaryCondition} is used as the initial condition, and the integration constants are set to $A=1$ and $B=3/2$. (We have set these dimensionless constants $A$ and $B$ to be of order one since the throat radius is normalized to be one.) 
The dashed and dotted lines represent the asymptotic behaviors. The lines $\tilde R \simeq \sqrt{8/n(n-1)}\tau + b$ and $\Phi \simeq \Phi(\infty)$ in the region $\tau \gg 1$ are approximated by using the values around $\tau\sim 40$ as $b\simeq \tilde R(40)-\sqrt{8/n(n-1)}\cdot 40 $ and $ \Phi(\infty)\simeq \Phi(40)$.
While they exhibit the expected behavior, the remaining equation \eqref{eq:hamiltonianconst} (the Hamiltonian constraint) is also satisfied, as illustrated in Figure~\ref{fig:04braneHconstraint}. More precisely, the Hamiltonian constraint is given by the right-hand-side of \eqref{eq:hamiltonianconst} multiplied by $e^{-2\phi}$ which we plot in Figure~\ref{fig:04braneHconstraint}.

To check the validity of the above numerical calculation, we perform the same calculation for the case of $n=2$, for which analytical solutions have already been provided by \cite{Horowitz:1991cd}\footnote{We also found analytical solutions for $n=9$. See Appendix \ref{app:n9sol}.}. We then compare the analytical and numerical results.

The extremal solution for $n=2$ is given by
\begin{align}
  \label{eq3_1}
  e^{-2(\Phi-\Phi_\infty)} &= 1 - \frac{\ell_0}{y}, \\
  \label{eq3_2}
  \diff s^2_{(n+1)} &= \frac{\diff y^2}{(1 - \ell_0/y)^2} + y^2\diff\Omega_2^2, 
\end{align}
where $\ell_0^2=\a'\sC/2$. 
To set the gauge $N=1$, let us consider the coordinate transformation 
\begin{equation}
  \label{eq:gtyr}
  \diff r = \frac{\diff y}{1 - \ell_0/y}.
\end{equation}
Integrating \eqref{eq:gtyr}, we obtain, 
\begin{equation}
  \tilde r = \tilde y+\log(\tilde y - 1) + C, \quad (\tilde y := y/\ell_0, ~\tilde r := r/\ell_0),
\end{equation}
where $C$ is an integration constant. Solving this equation for $y$, we obtain
\begin{align}
  \label{eq3_4}
  \tilde y &= 1 + W\left(e^{\tilde r-1-C}\right),\\
  \label{eq:asyR6}
    &\sim \tilde r - C - \log\left(\tilde r-1-C\right),\qquad\tilde r-1-C\gg 1
\end{align}
where $W(f)$ is the Lambert $W$ function, which is the inverse function of $f(x)=x e^x$.  
Then, the solutions can be rewritten in terms of $\tilde r$ as follows, 
\begin{align}
  \label{eq3_5}
  \Phi &= -\frac{1}{2}\log\left(\frac{W\left(e^{\tilde r-1-C}\right)}{1 + W\left(e^{\tilde r-1-C}\right)}\right) + \Phi_\infty, \\
  \label{eq3_6}
  \diff s^2_{(n+1)} /\ell_0^2 &= \diff \tilde r^2 + \tilde R^2\diff\Omega_2^2, \\
  \label{eq3_7}
  \Bigl(\tilde R &= 1 + W\left(e^{\tilde r-1-C}\right)\Bigr),
\end{align}
where $\Phi_\infty$ is the dilaton value at $r\to\infty$. The coordinate $\tau$ is related to $\tilde r$ by $\tilde r =2\tau$.

\begin{figure}[htb]
  \begin{center}
    $n=2$\\
    \begin{minipage}{0.48\columnwidth}
      \begin{center}
        \begin{tikzpicture}
          \node at (0,0) {\includegraphics[scale=0.74]{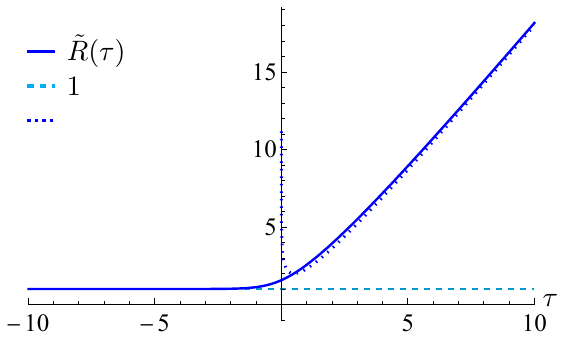}};
          \node at (-35/16,5/8) {\eqref{eq:asyR6}};
        \end{tikzpicture}
      \end{center}
    \end{minipage}
    \begin{minipage}{0.02\columnwidth}
      \hspace{0.02\columnwidth}
    \end{minipage}
    \begin{minipage}{0.48\columnwidth}
      \begin{center}
        \includegraphics[scale=0.74]{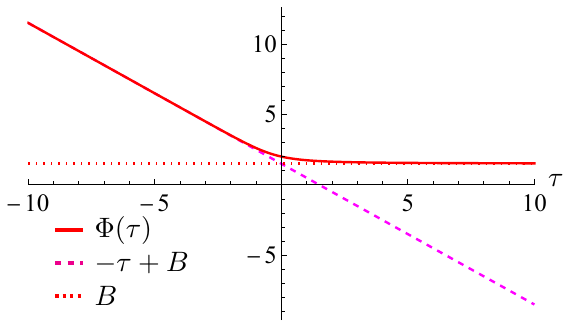}
      \end{center}
    \end{minipage}
  \end{center}
      \caption{The solutions for $n=2$. Left: $\tilde R(\tau)$. Right: $\Phi(\tau)$.}
  \label{fig:2branesol}
\end{figure}

\begin{figure}[htb]
  \begin{center}
    $n=2$ \\
    \includegraphics[scale=0.74]{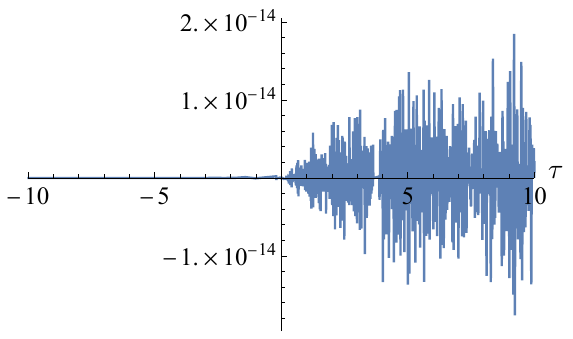}
    \caption{Hamiltonian constraint \eqref{eq:hamiltonianconst}$\times e^{-2\phi}$. These values are zero within the limits of numerical accuracy.}
    \label{fig:6braneHconst}
  \end{center}
\end{figure}

\begin{figure}[htb]
  \begin{center}
    \begin{minipage}{0.48\columnwidth}
      \begin{center}
        \includegraphics[scale=0.72]{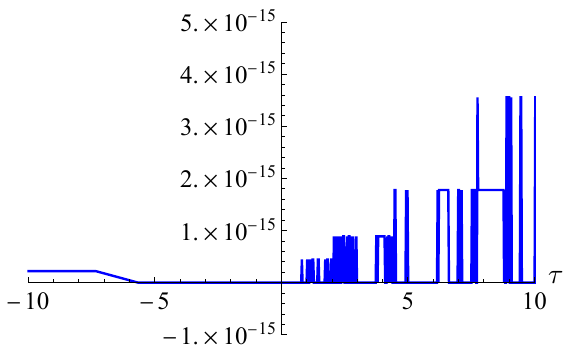}
      \end{center}
    \end{minipage}
    \begin{minipage}{0.02\columnwidth}
      \hspace{0.02\columnwidth}
    \end{minipage}
    \begin{minipage}{0.48\columnwidth}
      \begin{center}
        \includegraphics[scale=0.72]{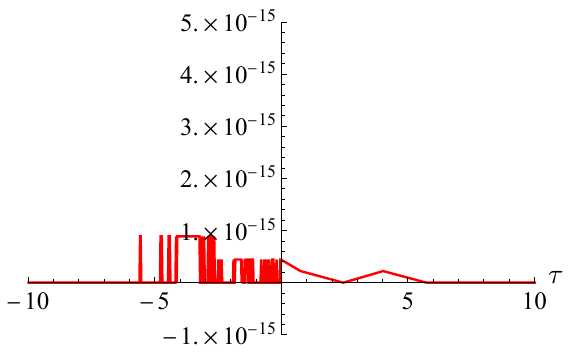}
      \end{center}
    \end{minipage}
  \end{center}
  \caption{Differences between the numerical and analytical solutions. Left: $|\tilde R_{\text{analytical}} - \tilde R_{\text{numerical}}|$. Right: $|\Phi_{\text{analytical}} - \Phi_{\text{numerical}}|$. These values are zero within the limits of numerical accuracy. }
  \label{fig:6branediff}
\end{figure}

Also, we can numerically solve the equations of motion for the case of $n=2$ in the gauge $N=1$, similar to $n=4,8$. Numerical solutions and the Hamiltonian constraint are illustrated in Figures \ref{fig:2branesol} and \ref{fig:6braneHconst}, respectively. The difference between the numerical and analytical solutions is illustrated in Figure~\ref{fig:6branediff}. 
In the analytical solution, we set $C=-1$ and $\Phi_\infty=3/2$. The numerical calculation accurately reproduces the analytical solutions. Therefore, we conclude that our numerical calculations for $n=4,8$ are also reliable.

\section{Non-extremal solutions} \label{sec:nonextremal}

In Section \ref{sec:extremal}, we considered the case in which all the directions $X^\alpha$ that are tangent to the brane are decoupled in the sense that they are free scalars on the worldsheet of the heterotic string theory. This case has the $\so(p,1)$ Lorentz symmetry and corresponds to the extremal solution. 

In this section we study non-extremal solutions. It is convenient to Wick-rotate the time direction so that the metric has Euclidean signature. We take the Euclidean time direction to be periodic, and we require that the solutions are smooth at the horizon where the circle $S^1$ for the time direction shrinks (see Fig. \ref{fig:non-extremal}).

\subsection{Basic facts}\label{sec:general}

First let us recall some basic facts about heterotic string theory. We assume that the $10$-dimensional metric is of the form
\beq
\d s^2_{(10)} = \d X^i \d X^i +  \d s^2_{(n+2)},
\eeq
where $X^i~(i=1,\cdots,p)$ are space coordinates parallel to the brane. The $(n+2)$-dimensional metric $\d s^2_{(n+2)}$ is taken to be of the form
\beq
\d s^2_{n+2} = \ell_0^2 e^{2\Sigma} \d t_E^2 + \d s^2_{n+1}
\eeq
where $\ell_0$ is defined by \eqref{eq:Sn-radius}, the $t_E$ is the Euclidean time direction which is taken to have the periodicity
\beq
t_E \sim t_E + 2\pi,
\eeq
and $\d s^2_{n+1}$ is independent of $t_E$.

The Ricci scalars $  \cR_{ G_{n+2} } $ and $  \cR_{G_{n+1}}$ for $\d s^2_{(n+2)}$ and $\d s^2_{(n+1)}$ are related by 
$\cR_{ G_{n+2} }  =   \cR_{ G_{n+1} } - 2e^{-\Sigma}\nabla^2_{G_{n+1}} e^\Sigma$.
Neglecting the directions $X^i$, the action becomes
\beq
& \int \d^{(n+2)} x  \sqrt{G_{n+2} } e^{-2\Phi_{n+2} }\left(\mathcal R_{G_{n+2}} + 4\left(\nabla_{G_{n+2}}\Phi_{n+2}   \right)^2 - \frac{\alpha'}{2}\tr' \left(|F|^2\right)\right) \nonumber \\
 & = 2\pi \ell_0 \int \d^{(n+1)} x  \sqrt{G_{n+1}} e^{-2\Phi_{n+1} }\left(\mathcal R_{G_{n+1}} + 4\left(\nabla_{G_{n+1}} \Phi_{n+1} \right)^2-(\nabla_{G_{n+1}} \Sigma)^2 - \frac{\alpha'}{2}\tr' \left(|F|^2\right)\right),
\eeq
where the two dilatons $\Phi_{n+2}$ and $\Phi_{n+1}$ are related by
\beq
\Phi_{n+1}=\Phi_{n+2}-\frac12 \Sigma. \label{eq:dilatonreduction}
\eeq
This is the well-known dimensional reduction in heterotic string theories.

In particular, the equation of motion for $\Sigma$ is given by 
\beq
\nabla_{G_{n+1}}^\mu(e^{-2\Phi_{n+1}} \nabla_{G_{n+1},\mu} \Sigma)=0. \label{eq:Sigmaeq}
\eeq
Thus, $\Sigma=0$ is a consistent solution. This justifies the ansatz \eqref{eq:extremalansatz} made in the extremal case. The same comment applies to the space directions $X^i$ tangent to the brane.

The $\d s_{(n+1)}^2$ is taken to be 
\beq
\diff s_{(n+1)}^2  =   N^2\diff r^2 + R^2\diff\Omega_n^2
\eeq 
as before.
The same calculation as in Section \ref{sec:extremal} which has led to \eqref{eq:extremalLagrangian} gives the Lagrangian as
\begin{align}
  \cL &= Ne^{-2\varphi}\left(-\frac{n}{8}\left(N^{-1}\sigma'\right)^2 - \frac{1}{8}\left(N^{-1}\Sigma'\right)^2 + \frac{1}{2}\left(N^{-1}\varphi' \right)^2 + e^{-2\sigma} - \frac{1}{2}e^{-4\sigma}\right), 
\end{align}
where we have used the change of variables \eqref{eq:variablechange}, except that $\phi$ is replaced by $\varphi$ which is given by
\begin{equation}
 \varphi =   \Phi - \frac{1}{2}\Sigma- \frac{n}{2}\sigma .
\end{equation}
Here, $ \Phi - \frac{1}{2}\Sigma$ is the effective dilaton in $(n+1)$-dimensions as can be seen in \eqref{eq:dilatonreduction}, and
$\varphi$ is the effective dilaton in one-dimension spanned by $r$. 
The Hamiltonian is given by
\begin{align}  
  \cH &= \Pi_\sigma\sigma' + \Pi_\Sigma\Sigma' + \Pi_\varphi\varphi'- \cL = NH, \\
  H &= e^{2\varphi}\left(-\frac{2}{n}\Pi_\sigma^2 - 2\Pi_\Sigma^2 + \frac{1}{2}\Pi_\varphi^2\right) + e^{-2\varphi}\left(-e^{-2\sigma} + \frac{1}{2}e^{-4\sigma}\right),
\end{align}
where the canonical momenta of $\Sigma$ is defined by
\begin{equation}
  \Pi_\Sigma = -\frac{1}{4}e^{-2\varphi}(N^{-1}\Sigma'),
\end{equation}
and the other canonical momenta are the same as \eqref{eq:CanMom}. The equations of motion are given by \eqref{eq:eoms} (with the replacement $\phi \to \varphi$) and 
\beq
N^{-1} \Sigma' = \frac{\partial H}{\partial \Pi_\Sigma},\quad N^{-1} \Pi'_\Sigma = - \frac{\partial H}{\partial \Sigma}.
 \label{eq:eomS}
\eeq

The metric and the dilaton are given by
\beq
\diff s_{(n+2)}^2 / \ell_0^2  &=   e^{2\Sigma} \d t_E^2 + \frac{8}{n(n-1)} N^2 \d \tau^2 + e^{2\sigma} \diff\Omega_n^2 \nonumber \\
e^{-2\Phi} & = e^{-2\varphi -\Sigma - n \sigma}. \label{eq:NonExtMetric}
\eeq
The equations of motion for $\sigma,\Sigma,\varphi$, and the Hamiltonian constraint in the gauge $N=1$ are given by
\begin{align}
  0 &= \frac{n}{4}(\sigma'' - 2\varphi'\sigma') - 2(e^{-2\sigma} - e^{-4\sigma}),\label{eq:nonexeomsigma} \\
  0 &= \Sigma'' - 2\varphi'\Sigma', \label{eq:nonexeomSigma}\\
  0 &= \varphi'' - \varphi'^2 - \frac{n}{4}\sigma'^2 - \frac{1}{4}\Sigma'^2 + 2e^{-2\sigma} - e^{-4\sigma}, \label{eq:nonexeomvarphi}\\
  0 &= \frac{n}{4}\sigma'^2 + \frac{1}{4}\Sigma'^2 - \varphi'^2 + 2e^{-2\sigma} - e^{-4\sigma}\label{eq:noneqhamiltonianconst}.
\end{align}
The equation \eqref{eq:nonexeomSigma} can be easily integrated to give
\beq
	\Sigma' = D e^{2\varphi}, \label{eq:Sigmaprime}
\eeq
where $D$ is a constant. The extremal solution is the case that $D=0$ so that $\Sigma$ is a constant.

\subsection{The behavior of solutions near horizon} \label{sec:Qnonexhorizon}

A horizon exists in non-extremal solutions. 
To find solutions, we use the standard condition that the Euclidean time direction $t_E$ shrinks at the horizon.
Recall that  the periodicity of $t_E$ has been taken to be $ 2\pi$.
The metric \eqref{eq:NonExtMetric} near the horizon which is smooth at the horizon must take the form,
\beq
\d s^2_{(n+2)}/\ell_0^2 \simeq \frac{8}{n(n-1)}  \left(\tau^2 \d t^2_E + d\tau^2\right) +\cdots, \label{eq:HorizonMetric}
\eeq
where the gauge $N=1$ is used and the position of the horizon is set to be  $\tau=0$. The $\tau$ and $t_E$ are the radial and angular directions in polar coordinates. This metric is smooth at $\tau=0$.

For the metric to be smooth, higher order terms are expanded in terms of $\tau^2$ (because $\tau$ itself is not smooth at $\tau=0$). Thus we take
\beq
e^{2\Sigma} &= {\frac{8}{n(n-1)}} \tau^2 (1+ A \tau^2+\cdots ),\\
e^{2\sigma} &= e^{2\sigma_0}(1+ B \tau^2+\cdots),\\
e^{-2\Phi} &= e^{-2\Phi_0}(1+ C \tau^2+\cdots).
\eeq
where $\sigma_0$ and $\Phi_0$ are the values of $\sigma$ and $\Phi$ at the horizon, respectively. By substituting these expansions into the equations \eqref{eq:nonexeomsigma}-\eqref{eq:Sigmaprime},
 the coefficients $A,B,C$ and $D$ are determined to be
 \beq
 A &= -\frac{2}{3}(2e^{-2\sigma_0}-e^{-4\sigma_0}),\\
 B &= \frac{4}{n}(e^{-2\sigma_0}-e^{-4\sigma_0}), \label{eq:Bvalue}\\
 C &= e^{-4\sigma_0},\\
D&= \sqrt{\frac{8}{n(n-1)}} e^{-2\Phi_0+n\sigma_0}. \label{eq:Dvalue}
\eeq
The solution near the horizon is given in terms of $A,B,C$  by
\beq
e^{-2\Phi} &=  e^{-2\Phi_0}(1+C \tau^2+\cdots),\label{eq:nearhorizonDilaton}\\
\d s^2_{n+2}/\ell_0^2 &= \frac{8}{n(n-1)}  \left[ \tau^2 (1+A\tau^2+\cdots) \d t^2_E + \d\tau^2 \right]  +  e^{2\sigma_0}(1+B \tau^2+\cdots) d\Omega_n^2.\label{eq:nearhorizonMetric}
\eeq
In addition to the parameter $\ell_0 = \sqrt{\alpha' \sC/n(n-1)}$ which is determined by the gauge flux as in \eqref{eq:Fsquared}, there are
two additional parameters $\sigma_0$ and $\Phi_0$. They are determined by the radius of $S^n$ and the value of the dilaton at the horizon. At spatial infinity $\tau \to \infty$, we have the vacuum expectation value of the dilaton and the ADM energy density of the brane. These two parameters at spatial infinity are related to the two parameters $\sigma_0$ and $\Phi_0$ at the horizon.

\subsection{The near extremal, near horizon limit} \label{sec:Qnonex}

One can find an analytical solution for the near extremal, near horizon limit in which the radius of $S^n$ is constant. Here we give this solution which is essentially the same as the cigar-like geometry discussed in \cite{Witten:1991yr}. 

If $\sigma_0=0$ at the horizon $\tau=0$, then the $B$ given by \eqref{eq:Bvalue} is zero. Moreover, by inspection of \eqref{eq:nonexeomsigma}, one can see that if $\sigma= 0$ and $\sigma'/\tau=0$ at $\tau=0$, then $\sigma$ actually remains to be 0 for all values of $\tau$. Thus, we can consider a solution with $\sigma=0$. 

Let us solve the equations \eqref{eq:nonexeomsigma}-\eqref{eq:Sigmaprime} when $\sigma = 0$. The equations in this case are simplified to
\beq
0&= \varphi'' -2\varphi'^2 +2,\label{eq:sigma01}\\
0&=\frac{D^2}{4}e^{4\varphi}-\varphi'^2+1 \label{eq:sigma02}.
\eeq
Unless $\varphi'=0$, the first equation follows from the second one by differentiating it and using it again. This is as expected, since the second equation is the Hamiltonian constraint which is basically ``the energy conservation'' (with zero energy), while the first equation is the equations of motion. Thus we can focus on the second equation. 

From the behavior $\varphi = \Phi-\frac12 \Sigma - \frac{n}{2} \sigma  \sim -\frac12 \log \tau$ near the horizon, we know that $\varphi' <0$ at least near the horizon. Thus we get
\beq
\varphi' = - \sqrt{\frac{D^2}{4}e^{4\varphi}+1}
\eeq
By setting $g =\frac{2}{D} e^{-2\varphi}$, the equation becomes $g'=2\sqrt{1+g^2}$. At the horizon $\tau=0$, we have $g(\tau=0)=0$. 
Then the solution is given by $g=\sinh (2\tau)$, or in other words 
\beq
e^{-2\varphi} = \frac{D}{2} \sinh (2\tau).
\eeq
From \eqref{eq:Sigmaprime}, we get $\Sigma' =2/ \sinh(2 \tau)$. By using $2/ \sinh(2 \tau) = ( \log \tanh \tau )'$, one can see that 
the solution is given by
\beq
e^{\Sigma} =  \sqrt{\frac{8}{n(n-1)} }   \tanh \tau,
\eeq
where we have taken into account the behavior \eqref{eq:nearhorizonMetric} near the horizon to determine the integration constant. Recalling that $e^{-2\Phi }=e^{-2\varphi - \Sigma}$, we get
\beq
e^{-2\Phi} = e^{-2\Phi_0} \cosh^2(\tau).
\eeq

In summary, the solution is given by
\beq
\d s^2_{n+2}/\ell_0^2
	&= \frac{8}{n(n-1)} \left( \tanh^2\tau \d t_E^2 + \d \tau^2 \right)+ 
 \d\Omega_n^2, \label{eq:NearExtremalMetric} \\
e^{-2\Phi} 
	&=e^{-2\Phi_0}  {\cosh^2\tau},\label{eq:NearExtremalDilaton}
\eeq
For $\tau \gg 1$, one can see that this reproduces the throat region of Section~\ref{sec:extremal}.

\subsection{The case of \texorpdfstring{$n=2$}{n=2}}\label{sec:non-ext:casetwodim}
For the 6-brane case (i.e. $n=2$), complete analytical solutions are known~\cite{Horowitz:1991cd}. They are given by
\beq
e^{-2\Phi } &=  e^{-2 \Phi_{\text{ext}} }\left(1-\frac{y_-}{y}\right),\label{eq:nonexanalyticdilaton2}\\
\d s^2_{n+2}/\ell_0^2 &= \frac{\left(1-\frac{y_+}{y}\right)}{\left(1-\frac{y_-}{y}\right)} \d t'^2_E  + \frac{\d y^2}{\left(1-\frac{y_+}{y}\right)\left(1-\frac{y_-}{y}\right)} + y^2 \d \Omega^2_2,\label{eq:nonexanalyticmetric2}
\eeq
where $y$ and $t'_E$ are coordinates, and $y_\pm$ and $\Phi_{\text{ext}}$ are constants. The $y_+$ and $y_-$ are related by
\beq
y_+y_- = 1 .
\eeq
The horizon is at $y=y_+ \geq 1$. The extremal limit is given by $y_+=y_-=1$.

The coordinate $y$ is related to the $\tau$ in \eqref{eq:NonExtMetric} in the gauge $N=1$ by
\beq
\tau &=\frac{1}{2} \int_{y_+}^y  \frac{\d y}{\sqrt{ \left(1-\frac{y_+}{y}\right)\left(1-\frac{y_-}{y}\right) }}  \nonumber \\
&= \frac{1}{2} \sqrt{(y-y_+)(y-y_-)} + \frac{y_+ +y_-}{2}  \log\left[ \frac{\sqrt{y-y_+}+\sqrt{y-y_-}}{\sqrt{y_+-y_-}}\right]. \label{eq:n=2rel1}
\eeq
By requiring that the metric is smooth at the horizon, one can check that $t'_E$ has periodicity $4\pi y_+$, and hence $t'_E$ is related to $t_E$ (which has periodicity $2\pi$) by
\beq
t'_E = 2y_+ t_E. \label{eq:n=2rel2}
\eeq
Then the functions $\Sigma$ and $\sigma$ are given by
\beq
e^{2\Sigma} = (2y_+)^2 \frac{\left(1-\frac{y_+}{y}\right)}{\left(1-\frac{y_-}{y}\right)}, \qquad e^{2\sigma}=y^2.\label{eq:n=2rel3}
\eeq
The parameters are related by
\beq
e^{-2\Phi_0} =  e^{-2 \Phi_{\text{ext}} }\left(1-\frac{y_-}{y_+}\right), \qquad e^{2\sigma_0}=y_+^2,\label{eq:n=2rel4}
\eeq
and $y_-=1/y_+$.

In particular, for the near extremal, near horizon limit, we take $y_+ - y_- \to 0$ (and hence $y_\pm \to 1$) and also $y-y_+ \to 0$ while fixing their ratio $(y-y_+)/(y_+-y_-)$ finite. Then,
\beq
\frac{y-y_+}{y_+ - y_-} \to \sinh^2\tau. 
\eeq
In this limit, we get
\beq
e^{-2\Phi } &= e^{-2\Phi_0 }{\cosh^2 \tau},\label{eq:n=2nearextp}\\
\d s^2_{n+2}/\ell_0^2 &=4 \tanh^2\tau \d t^2_E + 4 \d \tau^2 + d\Omega_2^2.\label{eq:n=2nearext}
\eeq
These are the same as \eqref{eq:nonexanalyticdilaton2} and \eqref{eq:nonexanalyticmetric2} for $n=2$.

 \subsection{Numerical solutions : the non-extremal case}\label{sec:non-extremal:sol}
 
\begin{figure}[h]
\centering
\begin{minipage}[b]{0.49\columnwidth}
    \centering
    $n=4$\\
    \includegraphics[width=0.9\columnwidth]{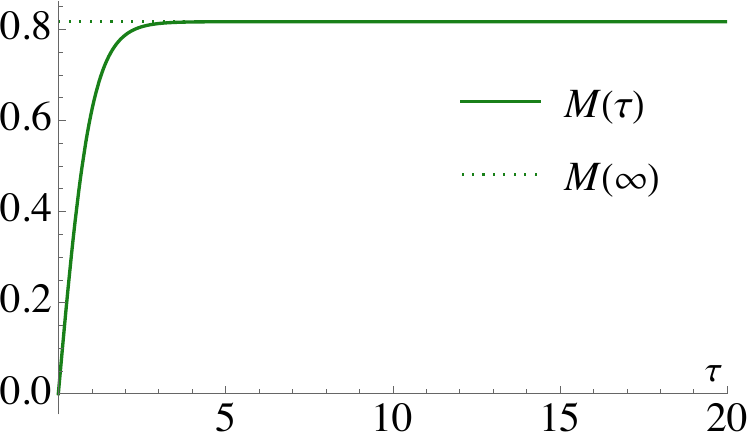}
\end{minipage}
\begin{minipage}[b]{0.49\columnwidth}
    \centering
    $n=8$\\
    \includegraphics[width=0.9\columnwidth]{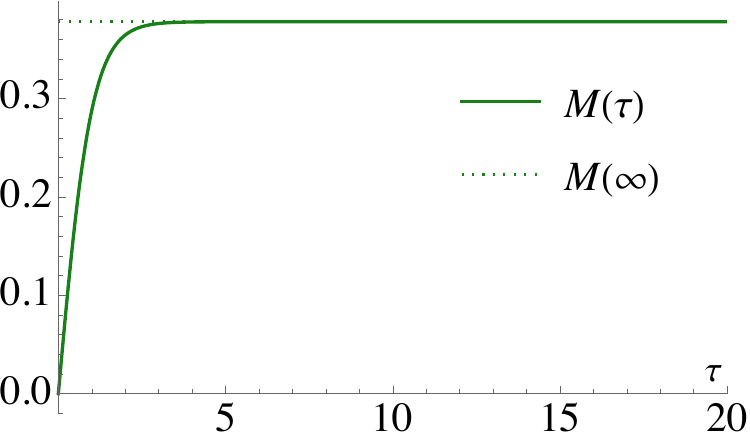}
\end{minipage} \\~\\
\begin{minipage}[b]{0.49\columnwidth}
    \centering
    \includegraphics[width=0.9\columnwidth]{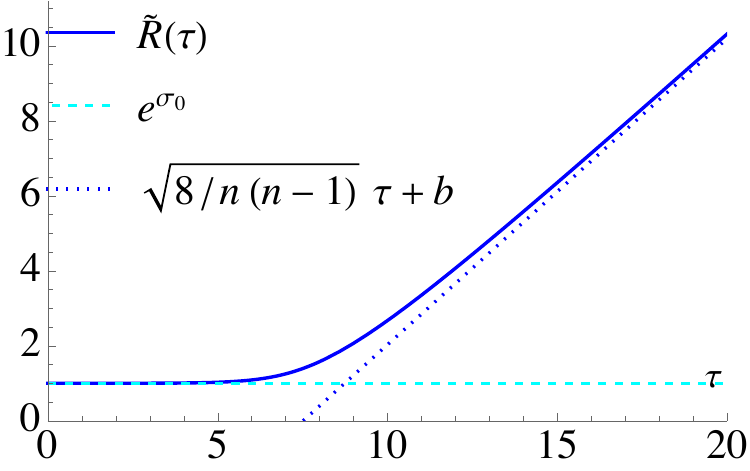}
\end{minipage}
\begin{minipage}[b]{0.49\columnwidth}
    \centering
    \includegraphics[width=0.9\columnwidth]{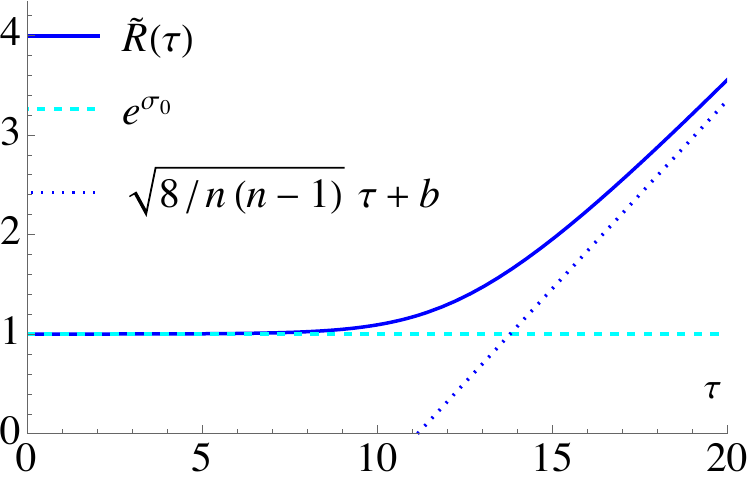}
\end{minipage} \\~\\
\begin{minipage}[b]{0.49\columnwidth}
    \centering
     \includegraphics[width=0.9\columnwidth]{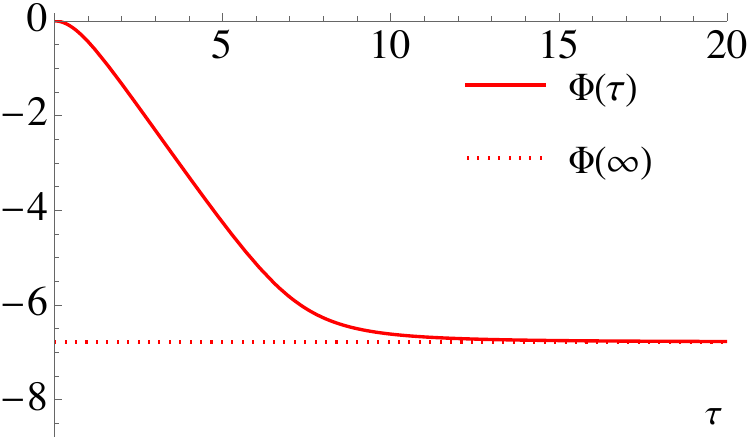}
\end{minipage}
\begin{minipage}[b]{0.49\columnwidth}
    \centering
     \includegraphics[width=0.9\columnwidth]{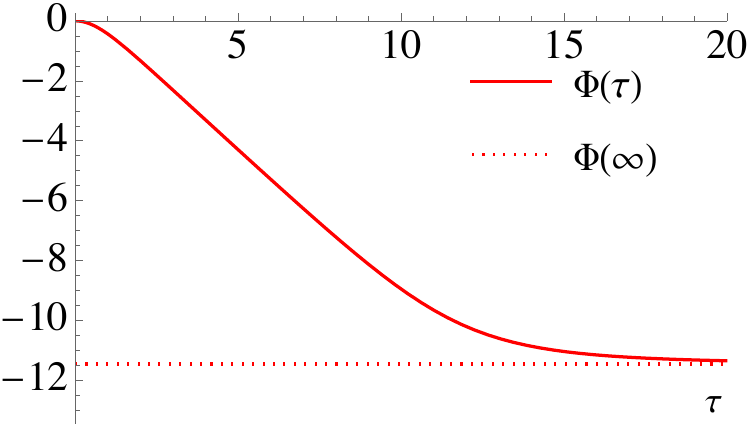}
\end{minipage}

\caption{Left: the solution with $n=4, \sigma_0=0.0001, \Phi_0=0$. Right: the solution with $n=8, \sigma_0=0.0001, \Phi_0=0$.}\label{fig:nonexn}
\end{figure}

 \begin{figure}[]
\centering
\begin{minipage}[b]{0.49\columnwidth}
    \centering
    \includegraphics[width=0.9\columnwidth]{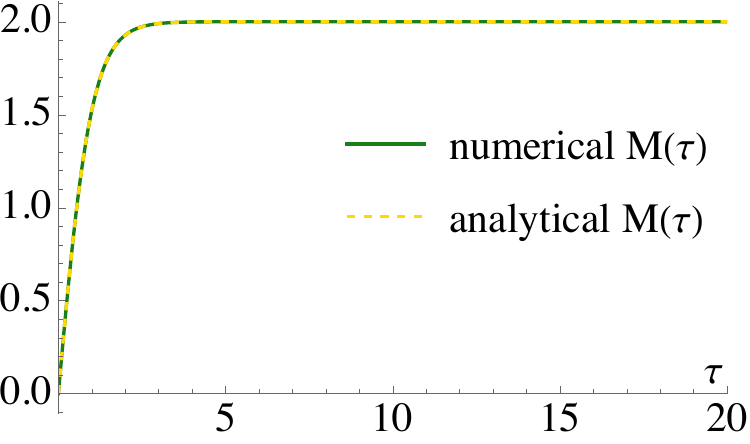}
\end{minipage} \\
\begin{minipage}[b]{0.49\columnwidth}
    \centering
    \includegraphics[width=0.9\columnwidth]{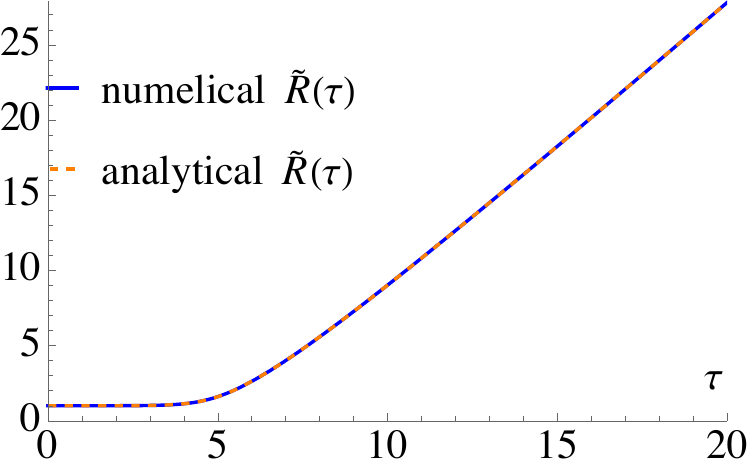}
\end{minipage}
\begin{minipage}[b]{0.49\columnwidth}
    \centering
     \includegraphics[width=0.9\columnwidth]{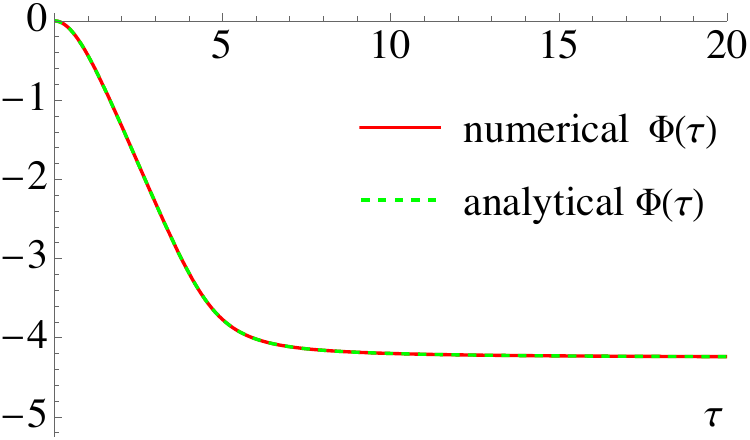}
\end{minipage}
\caption{The solution with $n=2, \sigma_0=0.0001, \Phi_0=0$.}\label{fig:nonexn=2}
\end{figure}

\begin{figure}[]
\centering
\begin{minipage}[b]{0.49\columnwidth}
    \centering
    \includegraphics[width=\columnwidth]{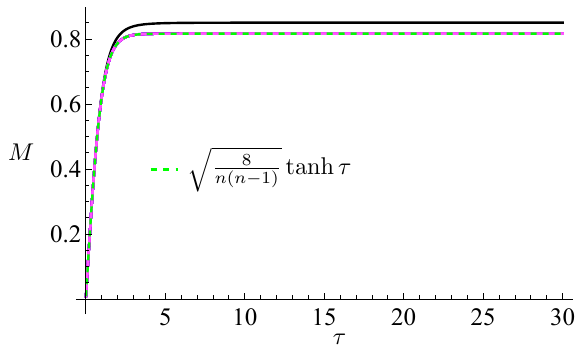}
\end{minipage}
\begin{minipage}[b]{0.49\columnwidth}
  \centering
    \includegraphics[scale=0.7]{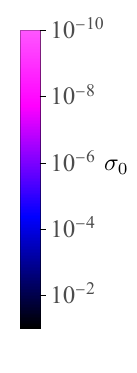}
\end{minipage}
\\
\begin{minipage}[t]{0.49\columnwidth}
    \centering
    \includegraphics[width=\columnwidth]{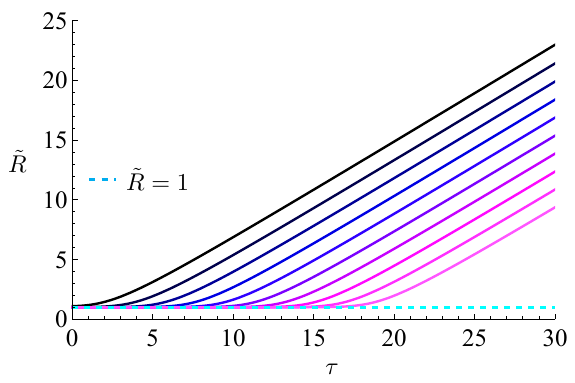}
\end{minipage}
\begin{minipage}[t]{0.49\columnwidth}
    \centering
    \includegraphics[width=\columnwidth]{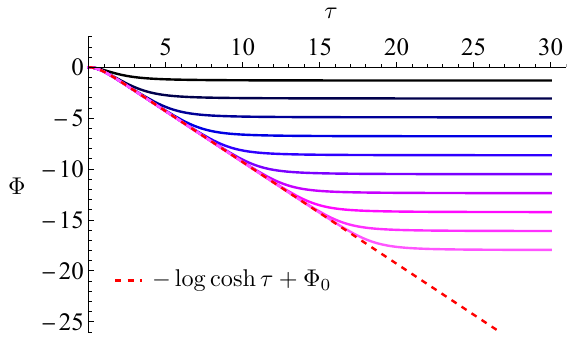}
\end{minipage}
\caption{The solution with $n=4, \Phi_0 =0$ and $ \sigma_0=10^{-1},10^{-2},\ldots,10^{-10}$. }\label{fig:nonexn=4-2}
\end{figure}

Now we present our numerical solutions. We solve \eqref{eq:nonexeomsigma}, \eqref{eq:nonexeomvarphi}, and \eqref{eq:Sigmaprime}. We plot the solutions for variables
\beq
M = e^{\Sigma}, \qquad \tilde R = e^{\sigma}, \qquad \Phi.
\eeq
The meaning of these variables are as follows. The $M$ is the radius of the $S^1$ in the Euclidean time direction in units of $\ell_0$. The $\tilde R$ is the radius of the $S^n$ in units of $\ell_0$. The $\Phi$ is the dilaton. 
We remark that the temperature $T$ of the black brane is given by
\beq
T = \frac{1}{2\pi \ell_0 M(\tau=\infty)}.
\eeq

The initial conditions are set by the behavior of the solutions near the horizon obtained in Section~\ref{sec:Qnonexhorizon}. More explicitly, we start from a very small but finite $\tau$, and set initial conditions at that value of $\tau$ by using \eqref{eq:nearhorizonMetric}. We have checked that the Hamiltonian constraint \eqref{eq:noneqhamiltonianconst} is satisfied to a good accuracy.

Numerical solutions for  $n=4 $, $\sigma_0=0.0001$, $ \Phi_0=0$ and $n=8$, $\sigma_0=0.0001$, $\Phi_0=0$ 
are shown in Figure~\ref{fig:nonexn}.
In the figures, $b$ and $\Phi(\infty)$ are constants determined by the behavior of $\tilde R$ and $\Phi$ at $\tau \to \infty$. More explicitly, we determined these constants by the behavior of the fields around $\tau \sim 40$. These solutions approach flat space solutions at $\tau \to \infty$ as expected.

When $n=2$, we have the analytical solution given by \eqref{eq:nonexanalyticdilaton2} and \eqref{eq:nonexanalyticmetric2}. The coordinates and parameters are related by \eqref{eq:n=2rel1}-\eqref{eq:n=2rel4}. In Figure~\ref{fig:nonexn=2}, one can see that the numerical and analytical solutions coincide with each other very well.

Next, we examine the behavior in the near-extremal limit \eqref{eq:NearExtremalDilaton}, \eqref{eq:NearExtremalMetric} as studied in subsection \ref{sec:Qnonex}. 
The value $\sigma_0=0$ is the near extremal, near horizon limit studied there.  
By setting $\sigma_0  \ll 1$, we expect to get a long throat region.
Numerical solutions with $n=4, \Phi_0 =0$, and $\sigma_0=10^{-k}$ for $k=1,2,\cdots,10$ are plotted in Figure~\ref{fig:nonexn=4-2}. In the figures, we have also plotted the analytical near-extremal solutoins \eqref{eq:NearExtremalMetric} and \eqref{eq:NearExtremalDilaton}. Indeed, we can see that the throat region appears which is well approximated by these equations.

\section{Conclusion}

In this paper, we constructed black 0- and 4-brane solutions for the non-supersymmetric heterotic branes proposed in \cite{Polchinski:2005bg,Bergshoeff:2006bs,Kaidi:2023tqo,Kaidi:2024cbx}. Specifically, we investigated both the extremal and non-extremal cases by numerical calculations. For the extremal case, the world sheet construction \cite{Kaidi:2023tqo,Kaidi:2024cbx} indicates that the solution is expected to have a throat region with a linear dilaton configuration in the near horizon limit. There should also be an asymptotically flat region. We have confirmed these properties. On the other hand, for the non-extremal (finite temperature) case in which the Euclidean time direction is compactified to $S^1$, the solution is expected to have a horizon where the $S^1$ radius shrinks to zero. In a limit, the geometry takes the form of a semi-infinite cigar \cite{Witten:1991yr}. Our solutions exhibit these expected properties. All these properties are similar to the case of NS5-branes~\cite{Horowitz:1991cd,Callan:1991at,Maldacena:1997cg}.

Furthermore, we checked the validity of our numerical calculations from several perspectives. First we checked that an analog of the ``Hamiltonian constraint'' (in the radial direction) is satisfied to a high degree of accuracy. We have also numerically reproduced the analytical solution of the 6-brane using the same methods as for the 0- and 4-branes. Moreover, in the near-extremal limit, the non-extremal solution goes to the extremal solution. Based on these results, we conclude that the desired black brane solutions have been obtained.

\section*{Acknowledgements}
We thank Kai Murai for helpful advice on Mathematica. The work of KY is supported in part by JST FOREST Program (Grant Number JPMJFR2030, Japan), 
MEXT-JSPS Grant-in-Aid for Transformative Research Areas (A) ”Extreme Universe” (No. 21H05188),
and JSPS KAKENHI (17K14265). The work of MF is supported by JST SPRING, Grant Number JPMJSP2114.

\appendix
\section{Examples of gauge field configurations }\label{sec:gaugefield}

In this appendix, we discuss some gauge field configurations, and how much the corresponding supergravity solutions are reliable.
If the scale of the radius of curvature of a solution is larger than the string scale $\sqrt{\alpha'}$, then $\alpha'$ corrections are small, and we can trust the supergravity action at the leading order of $\alpha'$. One such case is non-extremal solutions in which the size of the angular $S^n$ is large everywhere outsize of the horizon. Another case is extremal solutions when the parameter $\sC$ appearing in \eqref{eq:Fsquared} is large enough. This parameter is determined by gauge field configurations.

\subsection{The minimum gauge field configuration}

The minimum gauge field configurations discussed in \cite{Kaidi:2023tqo} are as follows. 
For the Lie algebra $\g = \so(32)$ or $\e_8 \times \e_8$, 
we have a subalgebra $\so(n) \subset \g$ for $n=4, 8$ that is given by:
\beq
\so(4) &\simeq \su(2) \times \su(2) \subset \e_8 \times \e_8, \\
\so(8)_s & \simeq\so(8)_v  \subset \so(32).
\eeq
Let us explain each of them. 

For $\so(4)$, the $\su(2) \subset \e_8$ is a subalgebra corresponding to a simple root. We take $\su(2)$'s in each of the two $\e_8$'s. The algebra $\su(2) \times \su(2)$ is of course isomorphic to the algebra $\so(4)$.

For $\so(8)$, the meanings of $\so(8)_s$ and $\so(8)_v$ are as follows. 
Both of them are just the $\so(8)$ algebra, but the isomorphism $\so(8)_s \simeq \so(8)_v $ uses an outer automorphism which exchanges the vector representation $v$ and one of the spinor representations $s$ of $\so(8)$. This is the famous automorphism which is part of the triality of $\so(8)$ (see e.g. \cite{Georgi:2000vve}.) The subalgebra $\so(8)_v \subset \so(32)$ is just taken as the obvious one in the vector representation. The triality has the property that  the quadratic Casimir invariants for the vector representation $v$ and the spinor representation $s$ are the same, and hence for an $\so(8)$ gauge field $F_{\so(8)}$ we have
\beq
\tr_v |F_{\so(8)}|^2 = \tr_s |F_{\so(8)}|^2.
\eeq

Now, suppose we have a subalgebra $\so(n) \subset \g$. On $S^n$, we can take a gauge field configuration such that the $\so(n)$ connection is the same as the Levi-Civita connection of the $S^n$ (which is assumed to have the standard round metric).
In an orthonormal frame, the Riemann curvature tensor of the sphere is given by
\beq
(\cR_{S^n})_{abcd} = R^{-2}(\delta_{ac} \delta_{bd} - \delta_{ad} \delta_{bc}) 
\eeq
where $a,b,c,d$ are indices in the orthonormal frame, and $R$ is the radius of $S^n$. By the above choice of the $\so(n)$ gauge field, the field strength $F_{\so(n)}$ is given by
\beq
(F_{\so(n)})_{abcd} = (\cR_{S^n})_{abcd}  ,
\eeq
where the first two indices $a,b$ are for gauge indices of $\so(n)$, and the last two indices $c,d$ are for the spacetime directions.
By taking this configuration for the subalgebra $\so(n) \subset \g$, we get
\beq
\tr' |F|^2 = \frac{1}{2} \cdot \frac{1}{2} \cdot (\cR_{S^n})_{abcd}  (\cR_{S^n})_{abcd} = \frac{n(n-1)}{2R^4}.
\eeq
In this case, the $\sC$ appearing in \eqref{eq:Fsquared} is given by $\sC = \frac12 n(n-1)$. 
It is known that these gauge field configurations are topologically nontrivial~\cite{Kaidi:2023tqo}.

\subsection{Configurations with large instanton numbers}\label{app:instanton}
In this and the next subsection, we focus on the case of the 4-brane, $n=4$.
In this case, the parameter $\sC$ is proportional to 
$|\nu|$, where $\nu$ and $-\nu$ are the instanton numbers of the two $E_8$ fields on the angular $S^4$, respectively.\footnote{
The two instanton numbers must be opposite to each other to satisfy the Bianchi identity $\d H \propto \tr' F_1^2+ \tr' F_2^2$, where $F_1$ and $F_2$ are the field strength 2-forms of the two $E_8$'s. 
}
Thus, the extremal solutions may be trusted if the instanton number $\nu$ is somewhat large. In this subsection, we discuss some examples of instanton configurations that can be constructed using $E_8$ and have large instanton numbers. 

For an $E_8$ gauge field, the instanton number is given by
\begin{equation}
  \label{eq:e8instanton}
  \nu:=\frac{1}{2}\int_{S^4}\tr'\left(\frac{\si F}{2\pi}\right)^2 = \frac{1}{120}\int_{S^4}\tr_{\rm adjoint}\left(\frac{\si F}{2\pi}\right)^2
\end{equation}
where $\tr'$ is defined by \eqref{eq:tr}.
The normalization is based on the following reason. A configuration with the 
smallest positive instanton number 
is a 1-instanton constructed using an $\su(2)$ subalgebra corresponding to a simple root of the gauge group. Such an $\su(2)$ is embedded in $\e_8$ as $\su(2) \subset \su(2) \times \e_7 \subset \e_8$. The adjoint representation ${\bf 248}$ of $\e_8$ decomposes under the $\su(2) \times  \e_7$ as
\beq
{\bf 248} \to ({\bf 3} \otimes {\bf 1}) \oplus ({\bf 2} \otimes {\bf 56}  ) \oplus ({\bf 1} \otimes {\bf 133}  ).
\eeq
For a single instanton of the  $\su(2) \subset \e_8$, one can check that $\nu=1$. 

For a given $\su(2)$ algebra, spherically symmetric solutions may only have the $\su(2)$ instanton number $\pm 1$~\cite{Yang:1977qv}. Therefore, to get a large instanton number of $\e_8$, let us consider other $\su(2)$ subalgebras within $\e_8$.

First, let us recall the relation between the trace in the $\su(2)$ spin $j=\frac{N-1}{2}$ representation and the spin $1/2$ representation. Let $T_a~(a=1,2,3)$ be the (anti-hermitian) basis of the $\su(2)$ algebra, 
\begin{equation}
  [\si T_a,\si T_b] = \si \epsilon_{abc}(\si T_c).
\end{equation}
For the spin $j=\frac{N-1}{2}$ representation, $\tr_{{\bf N}}(\si T_a\si T_b)= \frac12 T(N)\delta_{ab}$, where $T(N)$ is determined by setting $a=b=3$ as 
\begin{align}
  \nonumber
  T(N) &= 2\tr_{\bf N}(\si T_3\si T_3)
    = 2\sum_{m=-(N-1)/2}^{(N-1)/2} m^2, \\
    &= \frac{1}{6}N(N^2-1).
\end{align}
In particular, we have $T(2)=1$.
Thus, the trace in the representation ${\bf N}$ is $ \tr_{\bf N}(\si T_a\si T_b) =T(N) \tr_{\bf 2}(\si T_a\si T_b)$ and hence
we obtain
\begin{equation}
  \label{eq:jtohalf}
 \tr_{\bf N}(\si F)^2 = T(N) \tr_{\bf 2}(\si F)^2.
\end{equation}

Now, suppose that the adjoint representation ${\bf 248}$ of $\e_8$ decomposes under a subalgebra $\su(2) \subset \e_8$ as
\beq
  \e_8 \supset \su(2)~:~\mathbf{248} \to \bigoplus_{k} {\bf N_k}
\eeq
where ${\bf N_k}$ is the $N_k$-dimensional (spin $(N_k-1)/2$) representation of $\su(2)$. If the gauge field strength $F$ is in the subalgebra $\su(2)$, we get
\beq
 \tr_{\e_8,{\bf 248}} (\si F)^2 = \left( \sum_k T(N_k) \right)  \tr_{\su(2),{\bf 2}} (\si F)^2.
\eeq
Therefore, if we consider a single instanton for the $\su(2)$ subalgebra, the instanton number for $\e_8$ is given by
\beq
\nu = \frac{1}{60} \sum_k T(N_k). \label{eq:instformula}
\eeq

Let us discuss an example. There is a subalgebra under which the adjoint representation of $\e_8$ decomposes as (see e.g. \cite{Yamatsu:2015npn})
\begin{equation}
  \label{eq:branchSU2}
  \e_8 \supset \su(2)~:~\mathbf{248} \to \mathbf{3}\oplus\mathbf{15}\oplus\mathbf{23}\oplus\mathbf{27}\oplus\mathbf{35}\oplus\mathbf{39}\oplus\mathbf{47}\oplus\mathbf{59}.
\end{equation}
By using the formula \eqref{eq:instformula}, we get 
\beq
\nu &= \frac{1}{60} \left( T(3)+T(15)+T(23)+T(27)+T(35)+T(39)+T(47)+T(59)  \right) \nonumber \\
&=1240. \label{eq:1240}
\eeq
Examples using other subalgebras (see e.g. \cite{Yamatsu:2015npn}) are given in table \ref{tab:instantonnumbers}. 
\begin{table}[htb]
  \caption{The instanton numbers for the configurations using subalgebras $\su(2)$ of $\e_8$.}
  \label{tab:instantonnumbers}
  \begin{center}
    \begin{tabular}{c|l|c}
      subgroup & branching & instanton $\#$ \\\hline
      $\e_8\supset\su(2)$ & {\footnotesize$\mathbf{248} = \mathbf{3}\oplus\mathbf{15}\oplus\mathbf{23}\oplus\mathbf{27}\oplus\mathbf{35}\oplus\mathbf{39}\oplus\mathbf{47}\oplus\mathbf{59}$} & 1240 \\
      $\e_8\supset\su(2)$ & {\footnotesize$\mathbf{248} = \mathbf{3}\oplus\mathbf{11}\oplus\mathbf{15}\oplus\mathbf{19}\oplus\mathbf{23}\oplus\mathbf{27}\oplus\mathbf{29}\oplus\mathbf{35}\oplus\mathbf{39}\oplus \mathbf{47}$} & 760 \\
      $\e_8\supset\su(2)$ & {\footnotesize$\mathbf{248} = \mathbf{3}\oplus\mathbf{7}\oplus\mathbf{11}\oplus\mathbf{15}\oplus\mathbf{17}\oplus\mathbf{19}\oplus\mathbf{23}\oplus\mathbf{23}\oplus\mathbf{27}\oplus\mathbf{29}\oplus\mathbf{35}\oplus\mathbf{39}$} & 520 
    \end{tabular}
  \end{center}
\end{table}

\subsection{Central charge}

To get a sense of how large the instanton number should be for the supergravity solution to be reliable,
we consider the change in the central charge of the worldsheet sigma model with target space $S^n$. When the radius of $S^n$ is infinitely large, the (right-moving\footnote{We use the convention that the internal current algebra theory of the heterotic string theory is left-moving.}) central charge of $S^n$ is given by $c_R=\frac{3}{2} n$ which comes from just $n$ coordinates of $S^n$ and their fermionic superpartners. In the near horizon region, the radius $R$ of $S^n$ is given by a constant, $R=\ell_0$. Then the sigma model with target space $S^n$ receives quantum corrections which are power series of $\alpha'/R^2$. Suppose that the central charge at the radius $R=\ell_0$ is given by
\beq
c_R = \frac{3}{2}n - \delta c,
\eeq
where $\delta c$ is the quantum correction.
For the supergravity solution to be reliable, we need $\delta c \ll 1$. At least we can say that if $\delta c$ is too large such that $c_R<0$ or in other words $\delta c >\frac{3}{2}n$, then the solution is not reliable.

The correction $\delta c$ can be computed from the linear dilaton CFT. The total central charge is fixed in string theory, and hence the decrease $\delta c$ of the central charge of $S^n$ must be compensated by the increase in the linear dilaton CFT.

When the spacetime dilaton configuration is $\Phi(X) = V_\mu X^\mu$, the contribution to the central charge from the effect of the nontrivial dilaton configuration is given by
\begin{equation}
 \delta c = 6\alpha'V^2.
\end{equation}
In the throat region, the dilaton and metric are given by
\begin{align}
  \Phi &= -\frac{n(n-1)}{\sqrt{8\alpha'\sC}}r, \\
  \diff s^2 &= \diff X^\alpha\diff X_\alpha + \diff r^2 + \frac{\alpha'\sC}{n(n-1)}\diff\Omega_n^2.
\end{align}
Then the central charge is
\begin{equation}
\delta c = 6\alpha'\left(-\frac{n(n-1)}{\sqrt{8\alpha'\sC}}\right)^2 = \frac{3}{4}\frac{n^2(n-1)^2}{\sC}. \label{eq:deltac}
\end{equation}

Let us restrict our attention to the case of the 4-brane ($n=4$). Recall that $\sC$ is defined by~\eqref{eq:Fsquared}. We denote the field strength 2-forms of the two $E_8$ gauge fields as $F_1$ and $F_2$, respectively. Then, for the case of the 4-brane, \eqref{eq:Fsquared} is given by
\beq
\tr' |F_1|^2 + \tr' |F_2|^2 = \frac{\sC}{R^4},
\eeq
where $R$ is the radius of $S^n$, and $|F|^2 = \frac{1}{2}  F_{ij}^\dagger F^{ij}$. An (anti-)instanton configuration on $S^4$ satisfies the (anti-)self-dual equation $F =\pm \star F$, where $\star$ is the Hodge star on $S^4$. Thus
\beq
\int_{S^4} \tr' |F_k|^2 = \left| \int_{S^4} \tr' (\si F)^2 \right|  = 8\pi^2| \nu_k|, \qquad (k=1,2)
\eeq
where $\nu_k$ are the instanton numbers of the two $E_8$'s, given by $(\nu_1, \nu_2) = (\nu, -\nu)$.
On the other hand,
\beq
\int_{S^4} \frac{\sC}{R^4} = \frac{8\pi^2}{3} \sC.
\eeq
Therefore,
\beq
\sC = 3(|\nu_1| + |\nu_2|) = 6 |\nu|.
\eeq
Substituting it in \eqref{eq:deltac}, we get
\beq
\delta c =  \frac{18}{|\nu|}.
\eeq
This is the leading order result in an expansion in powers of $\alpha'/R^2= \alpha'/\ell_0^2 \sim 1/\sC$.

For instance, the instanton number \eqref{eq:1240} gives
\begin{equation}
  c_{\text{linear dilaton}} = \frac{18}{1240} = \frac{9}{620}.
\end{equation}
This is much smaller than $\frac32 n=6$.

\section{\texorpdfstring{$n=9$}{n=9} Solutions}\label{app:n9sol}
We have not found analytical solutions of the equations of motion \eqref{eq:eomsigma} and \eqref{eq:eomphi}  for general $n$, but we can solve these equations for $n=2$ and $n=9$. For $n=2$, the solution is equivalent to those in \cite{Horowitz:1991cd}. Here, we provide a solution for $n=9$. 

Eliminating $\phi'$ and $\sigma$ from the equation of motion \eqref{eq:eomphi} by using the Hamiltonian constraint \eqref{eq:hamiltonianconst}, we obtain,
\begin{equation}
  \phi'' = \frac{n}{2}\sigma'^2. \label{eq:phi''}
\end{equation}
We may change the radial coordinate from $\tau$ to $\sigma$, and regard $\phi' =\frac{\d \phi}{\d \tau}$ as a function of $\sigma$ instead of $\tau$. We denote this function as $f(\sigma)$, i.e.,  $\phi' = f(\sigma)$. Then from \eqref{eq:phi''} we obtain,
\begin{equation}
  \label{eq:sigmaprime}
  \sigma' = \frac{2}{n}\frac{\diff f}{\diff\sigma}.
\end{equation}
Substituting \eqref{eq:sigmaprime} and $\phi'=f$ into the Hamiltonian constraint \eqref{eq:hamiltonianconst}, we obtain,
\begin{equation}
  \label{eq:f}
  \frac{1}{n}\left(\frac{\diff f}{\diff \sigma}\right)^2 - f^2 + 2e^{-2\sigma} - e^{-4\sigma}=0.
\end{equation}
For $n=2$, 
one can check that $f(\sigma) = -2e^{-\sigma} + e^{-2\sigma}$ is a solution. It turns out to give the extremal 6-brane solution obtained in \cite{Horowitz:1991cd} after a short calculation.

Are there any other such solutions? Let us consider an ansatz $f(\sigma) = Ae^{-p\sigma} + Be^{-q\sigma}$, where $A,B,p,q$ are constants. Then we substitute it into \eqref{eq:f}. In addition to $n=2$, we can find the solution $f(\sigma) =\frac{1}{2}(-3e^{-\sigma}+e^{-3\sigma})$ for $n=9$. Then we get
\begin{align}
  \label{eq:sp9}
  \sigma' &= \frac{1}{3}(e^{-\sigma}-e^{-3\sigma}), \\
  \label{eq:pp9}
  \phi' &= \frac{1}{2}(-3e^{-\sigma}+e^{-3\sigma}).
\end{align}

Equations \eqref{eq:sp9} and \eqref{eq:pp9} are solvable. It is convenient to treat $\tilde R = e^{\sigma}$ as a variable when we solve these equations. In terms of $\tilde R$, equation \eqref{eq:pp9} becomes,
\begin{equation}
  \frac{\diff\phi}{\diff\tilde R} = \frac{1}{\tilde R} \left( \frac{\phi'}{\sigma'} \right) = -\frac{3}{2} \left(\frac{3\tilde R^2 - 1}{\tilde R^3 - \tilde R} \right).
\end{equation}
It is straightforward to integrate this equation, and we obtain,
\begin{equation}
  \phi = -\frac{3}{2}\log\left(\tilde R^3 - \tilde R\right) + \log g_s,
\end{equation}
where $\log g_s$ is an integration constant. In terms of $\Phi = \phi+ \frac{n}{2} \sigma$, this equation becomes,
\begin{equation}
  e^{-2\Phi} = g_s^{-2}\left(1 - \tilde R^{-2}\right)^3.
\end{equation}
On the other hand, using $ R$ as a variable, $\diff r$ becomes
\begin{equation}
  \diff r =  \sqrt{\frac{8}{n(n-1)}} \ell_0 \d \tau =  \frac{\ell_0}{3}  \frac{\diff \tilde R}{\diff e^{\sigma} /\diff \tau } = \ell_0  \frac{\diff \tilde R}{1 - \tilde R^{-2}} 
  = \frac{\diff R}{1 -  \left( \frac{\ell_0}{R} \right)^{2}   }  
\end{equation}
where we have used \eqref{eq:variablechange}. 
We obtain a solution for $n=9$ given by
\begin{align}
  e^{-2\Phi} &= g_s^{-2}\left(1 - \left( \frac{\ell_0}{R} \right)^{2}\right)^3, \\
  \diff s^2_{(10)} &= \left(\frac{\diff R}{1 -  \left( \frac{\ell_0}{R} \right)^{2}   }  \right)^2 + R^2\diff\Omega_9^2.
\end{align}

The physical interpretation this solution is not clear. In \cite{Alvarez-Garcia:2024vnr}, a $(-1)$-brane in $\Spin(32)/\bZ_2$ heterotic string theory has been investigated and we might at first hope that the solution is the $(-1)$-brane discussed there. Unfortunately, the above solution itself  has an infinite action and cannot give a finite contribution to physical processes.

\bibliographystyle{ytphys}
\bibliography{ref}

\providecommand{\href}[2]{#2}\begingroup\raggedright\begin{thebibliography}{10}

\bibitem{Polchinski:2003bq}
J.~Polchinski, {\slshape {Monopoles, duality, and string theory},} \href{http://dx.doi.org/10.1142/S0217751X0401866X}{{\em Int. J. Mod. Phys. A} {\bfseries 19S1} (2004) 145--156}, \href{http://arxiv.org/abs/hep-th/0304042}{{ arXiv:hep-th/0304042}}.

\bibitem{Banks:2010zn}
T.~Banks and N.~Seiberg, {\slshape {Symmetries and Strings in Field Theory and Gravity},} \href{http://dx.doi.org/10.1103/PhysRevD.83.084019}{{\em Phys. Rev. D} {\bfseries 83} (2011) 084019}, \href{http://arxiv.org/abs/1011.5120}{{ arXiv:1011.5120~[hep-th]}}.

\bibitem{Polchinski:1995mt}
J.~Polchinski, {\slshape {Dirichlet Branes and Ramond-Ramond charges},} \href{http://dx.doi.org/10.1103/PhysRevLett.75.4724}{{\em Phys. Rev. Lett.} {\bfseries 75} (1995) 4724--4727}, \href{http://arxiv.org/abs/hep-th/9510017}{{ arXiv:hep-th/9510017}}.

\bibitem{Misner:1957mt}
C.~W. Misner and J.~A. Wheeler, {\slshape {Classical physics as geometry: Gravitation, electromagnetism, unquantized charge, and mass as properties of curved empty space},} \href{http://dx.doi.org/10.1016/0003-4916(57)90049-0}{{\em Annals Phys.} {\bfseries 2} (1957) 525--603}.

\bibitem{Banks:1988yz}
T.~Banks and L.~J. Dixon, {\slshape {Constraints on String Vacua with Space-Time Supersymmetry},} \href{http://dx.doi.org/10.1016/0550-3213(88)90523-8}{{\em Nucl. Phys. B} {\bfseries 307} (1988) 93--108}.

\bibitem{Garfinkle:1990qj}
D.~Garfinkle, G.~T. Horowitz, and A.~Strominger, {\slshape {Charged black holes in string theory},} \href{http://dx.doi.org/10.1103/PhysRevD.43.3140}{{\em Phys. Rev. D} {\bfseries 43} (1991) 3140}. [Erratum: Phys.Rev.D 45, 3888 (1992)].

\bibitem{Harlow:2018jwu}
D.~Harlow and H.~Ooguri, {\slshape {Constraints on Symmetries from Holography},} \href{http://dx.doi.org/10.1103/PhysRevLett.122.191601}{{\em Phys. Rev. Lett.} {\bfseries 122} (2019) 191601}, \href{http://arxiv.org/abs/1810.05337}{{ arXiv:1810.05337~[hep-th]}}.

\bibitem{Harlow:2018tng}
D.~Harlow and H.~Ooguri, {\slshape {Symmetries in quantum field theory and quantum gravity},} \href{http://dx.doi.org/10.1007/s00220-021-04040-y}{{\em Commun. Math. Phys.} {\bfseries 383} (2021) 1669--1804}, \href{http://arxiv.org/abs/1810.05338}{{ arXiv:1810.05338~[hep-th]}}.

\bibitem{McNamara:2019rup}
J.~McNamara and C.~Vafa, {\slshape {Cobordism Classes and the Swampland},} \href{http://arxiv.org/abs/1909.10355}{{ arXiv:1909.10355~[hep-th]}}.

\bibitem{Kaidi:2024cbx}
J.~Kaidi, Y.~Tachikawa, and K.~Yonekura, {\slshape {On non-supersymmetric heterotic branes},} \href{http://arxiv.org/abs/2411.04344}{{ arXiv:2411.04344~[hep-th]}}.

\bibitem{Kaidi:2023tqo}
J.~Kaidi, K.~Ohmori, Y.~Tachikawa, and K.~Yonekura, {\slshape {Nonsupersymmetric Heterotic Branes},} \href{http://dx.doi.org/10.1103/PhysRevLett.131.121601}{{\em Phys. Rev. Lett.} {\bfseries 131} (2023) 121601}, \href{http://arxiv.org/abs/2303.17623}{{ arXiv:2303.17623~[hep-th]}}.

\bibitem{Hellerman:2006ff}
S.~Hellerman and I.~Swanson, {\slshape {Dimension-changing exact solutions of string theory},} \href{http://dx.doi.org/10.1088/1126-6708/2007/09/096}{{\em JHEP} {\bfseries 09} (2007) 096}, \href{http://arxiv.org/abs/hep-th/0612051}{{ arXiv:hep-th/0612051}}.

\bibitem{Hellerman:2007zz}
S.~Hellerman and I.~Swanson, {\slshape {A Stable vacuum of the tachyonic E(8) string},} \href{http://arxiv.org/abs/0710.1628}{{ arXiv:0710.1628~[hep-th]}}.

\bibitem{Kaidi:2020jla}
J.~Kaidi, {\slshape {Stable Vacua for Tachyonic Strings},} \href{http://dx.doi.org/10.1103/PhysRevD.103.106026}{{\em Phys. Rev. D} {\bfseries 103} (2021) 106026}, \href{http://arxiv.org/abs/2010.10521}{{ arXiv:2010.10521~[hep-th]}}.

\bibitem{BoyleSmith:2023xkd}
P.~Boyle~Smith, Y.-H. Lin, Y.~Tachikawa, and Y.~Zheng, {\slshape {Classification of chiral fermionic CFTs of central charge $\le$ 16},} \href{http://dx.doi.org/10.21468/SciPostPhys.16.2.058}{{\em SciPost Phys.} {\bfseries 16} (2024) 058}, \href{http://arxiv.org/abs/2303.16917}{{ arXiv:2303.16917~[hep-th]}}.

\bibitem{Fraiman:2023cpa}
B.~Fraiman, M.~Gra\~na, H.~Parra De~Freitas, and S.~Sethi, {\slshape {Non-Supersymmetric Heterotic Strings on a Circle},} \href{http://arxiv.org/abs/2307.13745}{{ arXiv:2307.13745~[hep-th]}}.

\bibitem{DeFreitas:2024yzr}
H.~P. De~Freitas, {\slshape {T-duality for non-critical heterotic strings},} \href{http://arxiv.org/abs/2407.12923}{{ arXiv:2407.12923~[hep-th]}}.

\bibitem{Yonekura:2024spl}
K.~Yonekura, {\slshape {Dualities among Neveu-Schwarz sector branes in string theory},} \href{http://arxiv.org/abs/2403.14933}{{ arXiv:2403.14933~[hep-th]}}.

\bibitem{Etheredge:2024amg}
M.~Etheredge, B.~Heidenreich, and T.~Rudelius, {\slshape {A Distance Conjecture for Branes},} \href{http://arxiv.org/abs/2407.20316}{{ arXiv:2407.20316~[hep-th]}}.

\bibitem{Hamada:2024cdd}
Y.~Hamada and A.~Ishige, {\slshape {Investigating 9d/8d non-supersymmetric branes and theories from supersymmetric heterotic strings},} \href{http://arxiv.org/abs/2409.04770}{{ arXiv:2409.04770~[hep-th]}}.

\bibitem{Polchinski:2005bg}
J.~Polchinski, {\slshape {Open heterotic strings},} \href{http://dx.doi.org/10.1088/1126-6708/2006/09/082}{{\em JHEP} {\bfseries 09} (2006) 082}, \href{http://arxiv.org/abs/hep-th/0510033}{{ arXiv:hep-th/0510033}}.

\bibitem{Bergshoeff:2006bs}
E.~A. Bergshoeff, G.~W. Gibbons, and P.~K. Townsend, {\slshape {Open M5-branes},} \href{http://dx.doi.org/10.1103/PhysRevLett.97.231601}{{\em Phys. Rev. Lett.} {\bfseries 97} (2006) 231601}, \href{http://arxiv.org/abs/hep-th/0607193}{{ arXiv:hep-th/0607193}}.

\bibitem{Basile:2023knk}
I.~Basile, A.~Debray, M.~Delgado, and M.~Montero, {\slshape {Global anomalies \& bordism of non-supersymmetric strings},} \href{http://dx.doi.org/10.1007/JHEP02(2024)092}{{\em JHEP} {\bfseries 02} (2024) 092}, \href{http://arxiv.org/abs/2310.06895}{{ arXiv:2310.06895~[hep-th]}}.

\bibitem{Debray:2023rlx}
A.~Debray, {\slshape {Bordism for the 2-group symmetries of the heterotic and CHL strings},} \href{http://arxiv.org/abs/2304.14764}{{ arXiv:2304.14764~[math.AT]}}.

\bibitem{Kneissl:2024zox}
C.~Kneissl, {\slshape {Spin cobordism and the gauge group of type I/heterotic string theory},} \href{http://arxiv.org/abs/2407.20333}{{ arXiv:2407.20333~[hep-th]}}.

\bibitem{Montero:2020icj}
M.~Montero and C.~Vafa, {\slshape {Cobordism Conjecture, Anomalies, and the String Lamppost Principle},} \href{http://dx.doi.org/10.1007/JHEP01(2021)063}{{\em JHEP} {\bfseries 01} (2021) 063}, \href{http://arxiv.org/abs/2008.11729}{{ arXiv:2008.11729~[hep-th]}}.

\bibitem{Blumenhagen:2021nmi}
R.~Blumenhagen and N.~Cribiori, {\slshape {Open-closed correspondence of K-theory and cobordism},} \href{http://dx.doi.org/10.1007/JHEP08(2022)037}{{\em JHEP} {\bfseries 08} (2022) 037}, \href{http://arxiv.org/abs/2112.07678}{{ arXiv:2112.07678~[hep-th]}}.

\bibitem{Blumenhagen:2022mqw}
R.~Blumenhagen, N.~Cribiori, C.~Kneissl, and A.~Makridou, {\slshape {Dynamical cobordism of a domain wall and its companion defect 7-brane},} \href{http://dx.doi.org/10.1007/JHEP08(2022)204}{{\em JHEP} {\bfseries 08} (2022) 204}, \href{http://arxiv.org/abs/2205.09782}{{ arXiv:2205.09782~[hep-th]}}.

\bibitem{Andriot:2022mri}
D.~Andriot, N.~Carqueville, and N.~Cribiori, {\slshape {Looking for structure in the cobordism conjecture},} \href{http://dx.doi.org/10.21468/SciPostPhys.13.3.071}{{\em SciPost Phys.} {\bfseries 13} (2022) 071}, \href{http://arxiv.org/abs/2204.00021}{{ arXiv:2204.00021~[hep-th]}}.

\bibitem{Velazquez:2022eco}
D.~M. Vel\'azquez, D.~De~Biasio, and D.~Lust, {\slshape {Cobordism, singularities and the Ricci flow conjecture},} \href{http://dx.doi.org/10.1007/JHEP01(2023)126}{{\em JHEP} {\bfseries 01} (2023) 126}, \href{http://arxiv.org/abs/2209.10297}{{ arXiv:2209.10297~[hep-th]}}.

\bibitem{Dierigl:2020lai}
M.~Dierigl and J.~J. Heckman, {\slshape {Swampland cobordism conjecture and non-Abelian duality groups},} \href{http://dx.doi.org/10.1103/PhysRevD.103.066006}{{\em Phys. Rev. D} {\bfseries 103} (2021) 066006}, \href{http://arxiv.org/abs/2012.00013}{{ arXiv:2012.00013~[hep-th]}}.

\bibitem{Dierigl:2022reg}
M.~Dierigl, J.~J. Heckman, M.~Montero, and E.~Torres, {\slshape {IIB string theory explored: Reflection 7-branes},} \href{http://dx.doi.org/10.1103/PhysRevD.107.086015}{{\em Phys. Rev. D} {\bfseries 107} (2023) 086015}, \href{http://arxiv.org/abs/2212.05077}{{ arXiv:2212.05077~[hep-th]}}.

\bibitem{Debray:2023yrs}
A.~Debray, M.~Dierigl, J.~J. Heckman, and M.~Montero, {\slshape {The Chronicles of IIBordia: Dualities, Bordisms, and the Swampland},} \href{http://arxiv.org/abs/2302.00007}{{ arXiv:2302.00007~[hep-th]}}.

\bibitem{Dierigl:2023jdp}
M.~Dierigl, J.~J. Heckman, M.~Montero, and E.~Torres, {\slshape {R7-branes as charge conjugation operators},} \href{http://dx.doi.org/10.1103/PhysRevD.109.046004}{{\em Phys. Rev. D} {\bfseries 109} (2024) 046004}, \href{http://arxiv.org/abs/2305.05689}{{ arXiv:2305.05689~[hep-th]}}.

\bibitem{Alvarez-Garcia:2024vnr}
R.~\'Alvarez-Garc\'ia, C.~Knei{\ss}l, J.~M. Leedom, and N.~Righi, {\slshape {Open Strings and Heterotic Instantons},} \href{http://arxiv.org/abs/2407.20319}{{ arXiv:2407.20319~[hep-th]}}.

\bibitem{Angius:2024pqk}
R.~Angius, A.~M. Uranga, and C.~Wang, {\slshape {End of the World Boundaries for Chiral Quantum Gravity Theories},} \href{http://arxiv.org/abs/2410.07322}{{ arXiv:2410.07322~[hep-th]}}.

\bibitem{Horowitz:1991cd}
G.~T. Horowitz and A.~Strominger, {\slshape {Black strings and P-branes},} \href{http://dx.doi.org/10.1016/0550-3213(91)90440-9}{{\em Nucl. Phys. B} {\bfseries 360} (1991) 197--209}.

\bibitem{Abbott:2008cd}
M.~C. Abbott and D.~A. Lowe, {\slshape {Six-Dimensional Yang Black Holes in Dilaton Gravity},} \href{http://dx.doi.org/10.1016/j.physletb.2008.04.064}{{\em Phys. Lett. B} {\bfseries 664} (2008) 214--218}, \href{http://arxiv.org/abs/0802.2976}{{ arXiv:0802.2976~[hep-th]}}.

\bibitem{Aharony:2019zsx}
O.~Aharony, M.~Evtikhiev, and A.~Feldman, {\slshape {Little String Theories on Curved Manifolds},} \href{http://dx.doi.org/10.1007/JHEP10(2019)180}{{\em JHEP} {\bfseries 10} (2019) 180}, \href{http://arxiv.org/abs/1908.02642}{{ arXiv:1908.02642~[hep-th]}}.

\bibitem{Witten:1991yr}
E.~Witten, {\slshape {On string theory and black holes},} \href{http://dx.doi.org/10.1103/PhysRevD.44.314}{{\em Phys. Rev. D} {\bfseries 44} (1991) 314--324}.

\bibitem{Callan:1991at}
C.~G. Callan, Jr., J.~A. Harvey, and A.~Strominger, {\slshape {Supersymmetric string solitons},} \href{http://arxiv.org/abs/hep-th/9112030}{{ arXiv:hep-th/9112030}}.

\bibitem{Maldacena:1997cg}
J.~M. Maldacena and A.~Strominger, {\slshape {Semiclassical decay of near extremal five-branes},} \href{http://dx.doi.org/10.1088/1126-6708/1997/12/008}{{\em JHEP} {\bfseries 12} (1997) 008}, \href{http://arxiv.org/abs/hep-th/9710014}{{ arXiv:hep-th/9710014}}.

\bibitem{Georgi:2000vve}
H.~Georgi, \href{http://dx.doi.org/10.1201/9780429499210}{{\em {Lie Algebras In Particle Physics : from Isospin To Unified Theories}}}.
\newblock Taylor \& Francis, Boca Raton, 2000.

\bibitem{Yang:1977qv}
C.~N. Yang, {\slshape {Generalization of Dirac's Monopole to SU(2) Gauge Fields},} \href{http://dx.doi.org/10.1063/1.523506}{{\em J. Math. Phys.} {\bfseries 19} (1978) 320}.

\bibitem{Yamatsu:2015npn}
N.~Yamatsu, {\slshape {Finite-Dimensional Lie Algebras and Their Representations for Unified Model Building},} \href{http://arxiv.org/abs/1511.08771}{{ arXiv:1511.08771~[hep-ph]}}.

\end{thebibliography}\endgroup

\end{document}